# Kernel Bounds for Path and Cycle Problems*

Hans L. Bodlaender†     Bart M. P. Jansen†     Stefan Kratsch†

June 19, 2018

**Abstract**

Connectivity problems like $k$-PATH and $k$-DISJOINT PATHS relate to many important milestones in parameterized complexity, namely the Graph Minors Project, color coding, and the recent development of techniques for obtaining kernelization lower bounds. This work explores the existence of polynomial kernels for various path and cycle problems, by considering nonstandard parameterizations. We show polynomial kernels when the parameters are a given vertex cover, a modulator to a cluster graph, or a (promised) max leaf number. We obtain lower bounds via cross-composition, e.g., for HAMILTONIAN CYCLE and related problems when parameterized by a modulator to an outerplanar graph.

## 1 Introduction

Connectivity problems such as $k$-PATH and $k$-DISJOINT PATHS play important theoretical and practical roles in the field of parameterized complexity. On the practical side, $k$-PATH [22, ND29] has applications in computational biology [29] where the involved parameter is fairly small, thus giving an excellent opportunity to apply parameterized algorithms to find optimal solutions. On the theoretical side, these problems have triggered the development of very powerful algorithmic techniques. The $k$-DISJOINT PATHS problem [28] lies at the heart of the Graph Minors Algorithm, and is the source of the *irrelevant-vertex* technique. The *color coding* technique of Alon et al. [1] to solve $k$-PATH has found a wide range of applications and extensions [27, 9], and new methods of solving $k$-PATH are still developing [3]. Despite the success stories of parameterized algorithms for these problems, the quest for polynomial kernels has resulted in mostly negative results. Indeed, the failure to find a polynomial kernel for $k$-PATH was one of the main motivations for the development of the kernelization lower-bound framework of Bodlaender et al. [5]. Using the framework it was shown that $k$-PATH does not admit a polynomial kernel unless NP $\subseteq$ coNP/poly, even when restricted to very specific graph classes such as planar cubic graphs. It did not take long before related connectivity problems such as $k$-DISJOINT PATHS [7], $k$-DISJOINT CYCLES [7], $k$-CONNECTED VERTEX COVER [17], and restricted variants of $k$-CONNECTED DOMINATING SET [14] were also shown not to admit polynomial kernels unless NP $\subseteq$ coNP/poly.

*This work was supported by the Netherlands Organization for Scientific Research (NWO), project "KERNELS: Combinatorial Analysis of Data Reduction".

†Utrecht University, P.O. Box 80.089, 3508 TB Utrecht, The Netherlands,
{h.l.bodlaender,b.m.p.jansen,s.kratsch}@uu.nl

Thus it seems that connectivity requirements in a problem form a barrier to polynomial kernelizability when it comes to the natural parameterization by solution size $k$. Driven by the desire to obtain useful preprocessing procedures for such problems, we may therefore investigate the kernelization complexity for non-standard parameters. Early work by Fellows et al. [20] shows that such a different perspective can yield polynomial kernels: they proved that HAMILTONIAN CYCLE parameterized by the max leaf number of the input graph $G$, i.e. the maximum number of leaves in a spanning tree for $G$, admits a linear-vertex kernel. In this work we study the existence of polynomial kernels for various structural parameters such as the max leaf number, the size of a vertex cover, and the vertex-deletion distance to simple graph classes such as cluster graphs and outerplanar graphs. Our results are as follows[1]:

1. We introduce a widely applicable technique based on matchings in bipartite graphs to show that the problems LONG CYCLE, its directed and path variants, DISJOINT PATHS, and DISJOINT CYCLES, admit kernels with $\mathcal{O}(|X|^2)$ vertices when parameterized by a vertex cover $X$.

2. We generalize these results to the stronger (i.e. smaller) parameter "vertex-deletion distance to a cluster graph". For LONG CYCLE and LONG PATH this requires the trick of encoding clique sizes in binary (in a weighted variant) and applying a Karp reduction to get "back" to the original problem. On the one hand, this has the drawback of not giving an explicit polynomial size bound. On the other hand, the binary encoding of clique sizes seems to be more efficient for subsequent computations. For DISJOINT CYCLES and DISJOINT PATHS the Karp reduction can be avoided, as the length of cycles or paths is not important.

3. Using the same binary encoding trick we get polynomial kernels for LONG CYCLE and LONG PATH parameterized by the max leaf number, generalizing the result of Fellows et al. [20] for HAMILTONIAN CYCLE. For DISJOINT CYCLES and DISJOINT PATHS the encoding trick is again not necessary.

4. We give contrasting kernelization lower bounds using the recently introduced technique of cross-composition [6]: (a) DIRECTED HAMILTONIAN CYCLE PARAMETERIZED BY A MODULATOR TO BI-PATHS does not admit a polynomial kernel unless NP $\subseteq$ coNP/poly, where the parameter measures the vertex-deletion distance to a digraph whose underlying undirected graph is a path, and (b) we modify the construction to prove that HAMILTONIAN CYCLE PARAMETERIZED BY A MODULATOR TO OUTERPLANAR GRAPHS does not admit a polynomial kernel; both results assuming NP $\not\subseteq$ coNP/poly.

5. We initiate the parameterized complexity study of finding paths respecting forbidden pairs [22, GT54] under various parameterizations. We obtain W[1]-hardness proofs, FPT algorithms, kernel lower bounds, and para-NP-completeness results.

**Related work.** Chen and Flum showed that MAXIMAL $k$-PATH is in FPT, and that MAXIMAL INDUCED $k$-PATH is W[2]-complete [10]. Joined by Müller they studied various forms of kernelization lower bounds, and showed amongst others that $k$-POINTED PATH does not admit a parameter non-increasing polynomial kernelization unless P = NP, and that $k$-PATH does not have a polynomial kernel on connected planar graphs unless NP $\subseteq$ coNP/poly [11].

---
[1] It is easy to see that HAMILTONIAN CYCLE and HAMILTONIAN PATH are special cases of LONG CYCLE and LONG PATH respectively. Hence, upper and lower bounds transfer in the obvious way between them.

**Organization.** Section 2 introduces the necessary notation regarding graphs and parameterized complexity, and introduces the main problems studied in this work. In Section 3 we show a useful property of bipartite matchings, which helps to obtain kernelization results. Section 4 contains the positive results, i.e., polynomial kernels, for the mentioned path and cycle problems when parameterized by a vertex cover (Section 4.1), the max leaf number (Section 4.2), or a modulator to a cluster graph (Section 4.3). In Section 5 we show the mentioned lower bound results for directed and undirected HAMILTONIAN CYCLE. Section 6 contains our results for path problems with forbidden pairs. We conclude in Section 7.

## 2 Preliminaries

**Graphs.** All graphs are finite and simple, unless indicated otherwise. An undirected graph $G$ has a vertex set $V(G)$ and an edge set $E(G) \subseteq \binom{V(G)}{2}$. A directed graph $D$ has a vertex set $V(D)$ and a set of directed arcs $A(D) \subseteq V(D)^2$. All paths are assumed to be simple. We say that a matching $M$ in a graph *covers* a set of vertices $U$, if each vertex in $U$ is endpoint of an edge in $M$. We use $[n]$ as a shorthand for the set $\{1, 2, \ldots, n\}$. If $X$ is a finite set then $\binom{X}{n}$ denotes the set of all size-$n$ subsets of $X$.

For a directed graph $D$ and vertex $v$ we write $N_D^+(v) := \{u \mid (v, u) \in A(D)\}$ for the out-neighbors, $N_D^-(v) := \{u \mid (u, v) \in A(D)\}$ for the set of in-neighbors, and $N_D(v) := N_D^+(v) \cup N_D^-(v)$ for the set of all neighbors. If $(u, v)$ is an arc of $D$ then $u$ is the *head* of the arc and $v$ is its *tail*. If $C \subseteq A(D)$ is a Hamiltonian cycle in a digraph containing the arcs $(x_i, x_{i+1})$ for $i \in [k]$ then we say that the vertices $x_1, \ldots, x_{k+1}$ appear consecutively on $C$, in that order. We also use the undirected variant of this notion which is defined analogously. For a set of vertices $X$ in a digraph $D$ we use $D - X$ to denote the digraph which results after deleting all vertices of $X$ and their incident arcs from $D$; the concept is defined analogously for undirected graphs.

The underlying undirected graph of a digraph $D$ is the result of disregarding the orientation of the arcs and eliminating parallel edges. Let BI-PATHS (for bi-orientations of paths) be the class of digraphs whose underlying undirected graph is a path. *Outerplanar graphs* are those graphs which can be drawn in the plane without crossings such that all the vertices lie on the outer face; such graphs have treewidth at most two. *Cluster graphs* are disjoint unions of cliques. For a graph class $\mathcal{G}$ and a vertex set $X \subseteq V(G)$ of a graph $G$ such that $G - X \in \mathcal{G}$, we say that $X$ is a *modulator* to the class $\mathcal{G}$.

**Parameterized complexity and kernelization.** A parameterized problem $Q$ is a subset of $\Sigma^* \times \mathbb{N}$, the second component being the *parameter* which expresses some structural measure of the input. A parameterized problem is (strongly uniformly) *fixed-parameter tractable* if there exists an algorithm to decide whether $(x, k) \in Q$ in time $f(k)|x|^{\mathcal{O}(1)}$ where $f$ is a computable function [18].

A *kernelization algorithm* (or *kernel*) for a parameterized problem $Q$ is a polynomial-time algorithm which transforms an instance $(x, k)$ into an equivalent instance $(x', k')$ such that $|x'|, k' \leq f(k)$ for some computable function $f$, which is the *size* of the kernel. If $f$ is a polynomial then this is a *polynomial kernel* [23].

**Cross-composition.** To prove our lower bounds we use the framework of cross-composition [6], which builds on earlier work by Bodlaender et al. [5], and Fortnow and Santhanam [21].

**Definition 1** (Polynomial equivalence relation [6])**.** An equivalence relation $\mathcal{R}$ on $\Sigma^*$ is called a *polynomial equivalence relation* if the following two conditions hold:

1. There is an algorithm that given two strings $x, y \in \Sigma^*$ decides whether $x$ and $y$ belong to the same equivalence class in $(|x| + |y|)^{\mathcal{O}(1)}$ time.

2. For any finite set $S \subseteq \Sigma^*$ the equivalence relation $\mathcal{R}$ partitions the elements of $S$ into at most $(\max_{x \in S} |x|)^{\mathcal{O}(1)}$ classes.

**Definition 2** (Cross-composition [6])**.** Let $L \subseteq \Sigma^*$ be a set and let $Q \subseteq \Sigma^* \times \mathbb{N}$ be a parameterized problem. We say that $L$ *cross-composes* into $Q$ if there is a polynomial equivalence relation $\mathcal{R}$ and an algorithm which, given $r$ strings $x_1, x_2, \ldots, x_r$ belonging to the same equivalence class of $\mathcal{R}$, computes an instance $(x^*, k^*) \in \Sigma^* \times \mathbb{N}$ in time polynomial in $\sum_{i=1}^{r} |x_i|$ such that:

1. $(x^*, k^*) \in Q \Leftrightarrow x_i \in L$ for some $1 \leq i \leq r$,

2. $k^*$ is bounded by a polynomial in $\max_{i=1}^{r} |x_i| + \log r$.

**Theorem 1** ([6])**.** *If some set $L \subseteq \Sigma^*$ is NP-hard under Karp reductions and $L$ cross-composes into the parameterized problem $Q$ then there is no polynomial kernel for $Q$ unless $NP \subseteq coNP/poly$.*

**Problems considered in this work.** The main focus of this work lies on nonstandard parameterizations of LONG PATH, LONG CYCLE, HAMILTONIAN PATH, HAMILTONIAN CYCLE, DISJOINT PATHS, and DISJOINT CYCLES. We briefly introduce the classical unparameterized versions; the first four a quite well known: Given a graph $G$ and an integer $k$, LONG PATH and LONG CYCLE ask for the existence of a path or cycle respectively containing at least $k$ vertices. Given a graph $G$, HAMILTONIAN PATH and HAMILTONIAN CYCLE ask for the existence of a path or cycle respectively which contains all vertices of $G$.

> DISJOINT PATHS
> **Input:** A graph $G$, an integer $k$, and a set of $k$ vertex pairs $\{(s_1, t_1), \ldots, (s_k, t_k)\}$.
> **Question:** Is there a set of $k$ vertex-disjoint paths, such that each pair $(s_i, t_i)$ is connected by exactly one of the paths?

> DISJOINT CYCLES
> **Input:** A graph $G$ and an integer $k$.
> **Question:** Does $G$ contain a set of $k$ vertex-disjoint cycles?

Most of the variants with nonstandard parameters considered in this work are given by requiring an extra set $X$ (a modulator) to be given in the input such that $G - X$ has some special property (like being an independent set). The parameter is then chosen as $\ell := |X|$. For formal reasons this would require an explicit inclusion of $\ell$ in the input, which we omit for succinctness.

Further variants of path problems with forbidden pairs are introduced in Section 6.

## 3 A property of maximum matchings in bipartite graphs

The following theorem simplifies the argumentation needed for reduction rules that are based on assigning private choices to some entities, e.g., vertices or edges, using an auxiliary bipartite graph, useful for example for various problems parameterized by a vertex cover.

**Theorem 2.** *Let $G = (X \cup Y, E)$ be a bipartite graph. Let $M \subseteq E(G)$ be a maximum matching in $G$. Let $X_M \subseteq X$ be the set of vertices in $X$ that are endpoint of an edge in $M$. Then, for each $Y' \subseteq Y$, if there exists a matching $M'$ in $G$ that covers $Y'$, then there exists a matching $M''$ in $G[X_M \cup Y]$ that covers $Y'$.*

*Proof.* Let $G$, $M$, and $X_M$ be as stated in the theorem. Suppose the theorem does not hold for $Y' \subseteq Y$, and let $M'$ be a matching in $G$ that covers $Y'$. Over all such matchings $M'$, take one that covers the largest number of vertices in $X_M$. By assumption $M'$ is not a matching in $G[X_M \cup Y]$, so there is a vertex $y_0 \in Y'$ that is matched in $M'$ to a vertex in $X \setminus X_M$, say $x_0$. We use an iterative process to derive a contradiction, maintaining the following invariants:

- $x_0 \notin X_M$.
- $\{x_j, y_j\} \in M'$ for $0 \leq j \leq i$.
- $\{y_j, x_{j+1}\} \in M$ for $0 \leq j < i$.
- The vertices $x_j$ for $0 \leq j \leq i$ are distinct members of $X$, and vertices $y_j$ for $0 \leq j \leq i$ are distinct members of $Y$.

It is easy to verify that given our choice of $x_0, y_0$ these invariants are initially satisfied for $i = 0$. We now continue the process based on a case distinction:

1. If $y_i$ is not matched under $M$, then the sequence $(x_0, y_0, \ldots, x_i, y_i)$ is an $M$-augmenting path in $G$ since $x_0$ and $y_i$ are not matched under $M$, and all edges $\{y_j, x_{j+1}\}$ for $0 \leq j < i$ are contained in $M$. Hence $M'' := M \setminus \{\{y_j, x_{j+1}\} \mid 0 \leq j < i\} \cup \{\{x_j, y_j\} \mid 0 \leq j \leq i\}$ is a matching in $G$ larger than $M$, contradicting that $M$ is maximum.

2. In the remaining cases we may assume $y_i$ is matched under $M$, say $\{y_i, x_{i+1}\} \in M$. If there is an index $0 \leq j \leq i$ such that $x_{i+1} = x_j$ then $j > 0$ (since $x_0 \notin X_M$) and the edges $\{y_i, x_{i+1}\}$ and $\{y_{j-1}, x_j\}$ are both contained in $M$ and are distinct edges since $y_{j-1} \neq y_i$, contradicting the fact that $M$ is a matching. Hence $x_{i+1}$ is distinct from $x_j$ for $0 \leq j \leq i$.

3. If $x_{i+1}$ is not covered by $M'$ then the matching $M'' := M' \setminus \{\{x_j, y_j\} \mid 0 \leq j \leq i\} \cup \{\{y_j, x_{j+1}\} \mid 0 \leq j \leq i\}$ contains as many edges as $M'$ but covers more vertices of $X_M$, contradicting the choice of $M'$. Hence $x_{i+1}$ is covered by $M'$, say $\{x_{i+1}, y_{i+1}\} \in M'$. If there is an index $0 \leq j \leq i$ such that $y_{i+1} = y_j$ then $\{x_{i+1}, y_{i+1}\}$ and $\{x_j, y_j\}$ are two distinct edges in $M'$ incident on $y_{i+1}$, contradicting that $M'$ is a matching. Hence $y_{i+1}$ is distinct from $y_j$ for $0 \leq j \leq i$. Now observe that the invariant holds for $i+1$, and we may proceed with the next step of the process.

By the last property of the invariant, the process must end. Hence the assumption that there is no matching in $G[X_M \cup Y]$ which covers $Y'$ leads to a contradiction, which concludes the proof. $\square$

## 4 Polynomial kernels for path and cycle problems

This section provides polynomial kernels for various path and cycle problems when parameterized by a vertex cover (Section 4.1), the max leaf number (Section 4.2), or a modulator to a cluster graph (Section 4.3).

## 4.1 Parameterization by a vertex cover

In this section we consider path and cycle problems when parameterized by the size $\ell$ of a given vertex cover, focusing mainly on the LONG CYCLE problem, for which we present a kernel with $\mathcal{O}(\ell^2)$ vertices.

> LONG CYCLE PARAMETERIZED BY A VERTEX COVER
> **Input:** A graph $G$, an integer $k$, and a set $X \subseteq V(G)$ such that $X$ is a vertex cover.
> **Parameter:** $\ell := |X|$.
> **Question:** Does $G$ contain a cycle of length at least $k$?

We need only one reduction rule to get a kernelization, it uses a bipartite *connection graph* $H = H(G, k, X)$: One color class consists of the vertices in the independent set $I = V(G) \setminus X$, and the other consists of all (unordered) pairs of distinct vertices in $X$. We take an edge from a vertex $v \in I$ to a vertex representing the pair $\{p, q\} \subseteq X$, if and only if $v$ is adjacent to $p$ and to $q$.

**Reduction Rule 1.** Given $(G, k, X)$, if $k \leq 4$ then solve the problem (e.g. by the trivial $\mathcal{O}(n^4)$ algorithm) and return an equivalent dummy instance. Otherwise, construct the connection graph $H = H(G, k, X)$. Let $M$ be a maximum matching in $H$. Let $J \subseteq I$ be the vertices covered by an edge in $M$. Remove all vertices in $I \setminus J$ and their incident edges from $G$. Let $G'$ be the resulting graph, and return the instance $(G', k, X)$.

**Observation 1.** In Rule 1, $|J|$ is at most the number of pairs of distinct vertices in $X$, and hence after applying the rule, $G'$ has at most $\ell + \binom{\ell}{2} \in \mathcal{O}(\ell^2)$ vertices.

Correctness of the rule follows from the following lemma.

**Lemma 1.** *Let $(G, k, X)$ be an instance of* LONG CYCLE PARAMETERIZED BY A VERTEX COVER, *and let $(G', k, X)$ be the instance returned by Rule 1. Then $G$ has a cycle of length at least $k$ if and only if $G'$ has a cycle of length at least $k$.*

*Proof.* If $k \leq 4$ then the lemma holds trivially. Otherwise, we have that $G'$ is an induced subgraph of $G$ so cycles (in particular those of length at least $k$) in $G'$ exist also in $G$. It remains to look at the converse.

Let $C$ be a cycle of length at least $k \geq 5$ in $G$. Clearly, as $I = V \setminus X$ is an independent set, any vertices of $I$ which are in $C$ must be neighbored by vertices of $X$ on $C$. Let $v_1, \ldots, v_r$ be all vertices of $I$ contained in $C$ and let $p_i$ and $q_i$ be the predecessor and successor of $v_i$ on $C$, respectively (clearly $r \leq \ell$ but there might be far fewer vertices of $I$ on $C$). Since $C$ has length at least 5, it follows that $\{p_i, q_i\} \neq \{p_j, q_j\}$ for all $i, j \in [r]$ with $i \neq j$ (else it would have length 4). To show that $G'$ contains a cycle of length at least $k$, it suffices to find replacements for all vertices $v_i$ which are not in $J$ (and hence not in $G'$); for this we will use the matching.

Clearly, in $H = H(G, k, X)$ there is a matching $M$ covering $W := \{\{p_1, q_1\}, \ldots, \{p_r, q_r\}\}$, namely matching each pair to the corresponding vertex $v_i$. Further, by Rule 1, $J$ is the set of endpoints in $I$ of some maximum matching of $H$. Hence, by Theorem 2, there is a matching $M'$ covering $W$ in $H[J \cup W]$.

Let $v'_i$ denote the vertex matched to $\{p_i, q_i\}$ by $M'$, for $i \in \{1, \ldots, r\}$. It is easy to see that we may replace each $v_i$ on $C$ by $v'_i$ since $v'_i$ is adjacent to $p_i$ and $q_i$ in $G$, obtaining a cycle $C'$ which intersects $I$ only in vertices of $J$. Also, as all pairs $\{p_i, q_i\}$ are different, no vertex $v'_i$ is required twice. Hence, $C'$ is also a cycle of $G'$, and of length at least $k$. □

The kernelization result now follows from Lemma 1 and Observation 1, and noting that Rule 1 can be easily performed in polynomial time.

**Theorem 3.** LONG CYCLE PARAMETERIZED BY A VERTEX COVER *has a kernel with $\mathcal{O}(\ell^2)$ vertices.*

We note that in the obtained kernel the number of edges may still be cubic in $\ell$, giving an overall size bound of $\mathcal{O}(\ell^3)$ by using an adjacency matrix encoding. It would be interesting to know whether the number of edges can be reduced to $\mathcal{O}(\ell^2)$ and whether a size bound of $\mathcal{O}(\ell^2)$ could be showed to be tight, using recent results on polynomial lower bounds for kernelization [16, 15, 25].

**Further problems parameterized by vertex cover.** The same technique can be used for a number of additional problems, all parameterized by the size of a vertex cover. The basic argument is the same; the matching approach allows us to reroute any paths or cycles such that they use only matched vertices.

**Corollary 1.** LONG PATH, DISJOINT PATHS, *and* DISJOINT CYCLES *admit polynomial kernels with $\mathcal{O}(\ell^2)$ vertices.*

*Proof.* We briefly sketch how to modify the proof given for LONG CYCLE (Theorem 3).

For an instance $(G, k, X)$ of the LONG PATH problem, it is most convenient to introduce a universal vertex $v$ (adjacent to all vertices of $G$) to obtain an equivalent instance $(G', k+1, X \cup \{v\})$ of LONG CYCLE parameterized by a vertex cover. Each $k$-path in $G$ then corresponds to a $k+1$-cycle in $G'$ by adding $v$, and each $k+1$-cycle in $G'$ contains at least one $k$-path in $G$. We then apply the kernelization as for LONG CYCLE, and finish up by removing $v$ from the obtained instance.

For DISJOINT PATHS it is clear that each path fulfilling a request $(s_i, t_i)$ must contain at least one vertex of $X$, hence YES-instances have at most $\ell = |X|$ request pairs. Thus we may assume that all vertices of request pairs are contained in $X$ (else adding them will at most triple the size of $X$). Since the paths have to be vertex-disjoint, no two request pairs share any vertices (though the proof could be adapted to internally vertex-disjoint paths, and also to asking for multiple paths between certain pairs). It is hence clear that edges between vertices of different pairs cannot be part of any path. Therefore, the easiest way to argue correctness is to add all those edges to $G$, and observe that the kernelization for LONG CYCLE also preserves the existence of a cycle which traverses all request pairs in order, i.e., $(\ldots, s_1, \ldots, t_1, s_2, \ldots, t_2, s_3, \ldots)$. Such a cycle exists if and only if the $k$ requested disjoint paths exist and the instance is YES.

For the DISJOINT CYCLES problem we may proceed essentially as for LONG CYCLE, since each single cycle will be preserved (as the argument only comes down to providing the private shared neighbors). The only difference is that we also have to preserve cycles of length four, which may require two vertices of the independent set to be assigned (matched) to one pair of vertices of the vertex cover $X$. (We also must preserve cycles of length three of course, but only length four cycles may have the particular mentioned layout.) To do so create the auxiliary bipartite graph $H = H(G, k, X)$ but duplicate all vertices corresponding to (unordered) pairs of vertices from $X$. This way, each pair can receive two private shared neighbors. □

For HAMILTONIAN PATH and HAMILTONIAN CYCLE it is easy to see that any vertex cover of a YES-instance must have size at least least $\lfloor \frac{|V|}{2} \rfloor$, since vertices of the remaining independent set cannot be adjacent on a Hamiltonian path or cycle. Thus all nontrivial instances $(G, X)$ have $|V(G)| \leq 2|X| + 1 = \mathcal{O}(\ell)$.

Furthermore, the matching argument can also be applied to the directed versions of these problems. For this, match to each ordered pair $(u, v)$ with $u$ and $v$ being vertices of the provided vertex cover a vertex $p$ such that there are (directed) edges $(u, p)$ and $(p, v)$ (note that there will be a vertex $q$ for $(v, u)$ as well, with edges $(v, q)$ and $(q, u)$). This way, any directed paths or cycles can be rerouted to use only matched vertices from $V(G) \setminus X$, providing kernels of the same size (up to constant factors). Again, for HAMILTONIAN PATH and HAMILTONIAN CYCLE the simpler argument from the previous paragraph applies.

## 4.2 Parameterization by max leaf number

In this section we consider path and cycle problems parameterized by the max leaf number, i.e., the maximum number of leaves in any spanning tree of the graph. Deviating slightly from the standard use, we will take the max leaf number of a disconnected graph to be the sum of max leaf numbers taken over all connected components (alternatively one may restrict the question to connected graphs). We will use LONG CYCLE as a running example, but as in Section 4.1 it is easy to generalize the arguments to further problems. As the max leaf number of a graph cannot be verified in polynomial time, we consider the parameterization in the sense of a promise problem, e.g.:

> LONG CYCLE PARAMETERIZED BY MAX LEAF NUMBER (LCML)
> **Input:** A graph $G$ and two integers $k$ and $\ell$.
> **Parameter:** $\ell$.
> **Question:** If $G$ has max leaf number at most $\ell$, then decide whether $G$ contains a cycle of length at least $k$. Else the output may be arbitrary.

It is well known that a large graph having small max leaf number must contain long paths of degree two vertices and few vertices of degree at least three. The following bound was obtained by Fellows et al. [20] based on work by Kleitman and West [26]; it can be easily seen to hold for each connected component.

**Lemma 2** ([20]). *If a graph $G$ has max leaf number at most $\ell$ then it is a subdivision of some graph $H$ of at most $4\ell - 2$ vertices. In particular, $G$ has at most $4\ell - 2$ vertices of degree at least three.*

It is not hard to devise an FPT-algorithm for LCML.

**Lemma 3.** LONG CYCLE PARAMETERIZED BY MAX LEAF NUMBER *can be solved in time $2^{\mathcal{O}(\ell)} n^c$.*

*Proof.* Let $(G, k, \ell)$ be an instance of LCML. Let $B$ denote those vertices that have degree at least three in $G$. If $|B| > 4\ell - 2$ then by Lemma 2, the promise is not fulfilled as the max leaf number of $G$ is greater than $\ell$, we return NO. Otherwise, we have $|B| \leq 4\ell - 2$.

Now, replacing each path connecting two $B$ vertices (without visiting other $B$ vertices) by an edge with weight equal to the length of the path we obtain a multigraph with the same maximum cycle length. Clearly, a longest cycle will either consist of just two vertices, or use at most one of the edges between any two vertices. Hence, after checking whether there is a sufficiently long 2-vertex cycle, we may discard all but the longest edge connecting any two vertices. The remaining problem can be solved by Held-Karp style dynamic programming [24] on the remaining weighted graph with $|B| = \mathcal{O}(\ell)$ vertices; hence in time $2^{\mathcal{O}(\ell)} n^c$. □

The main idea for the kernelization is that one of two good cases must hold: Either all the path lengths are small enough such that a binary encoding of their length has size polynomial in $\ell$, or the total number $n$ of vertices is large enough such that the $2^{\mathcal{O}(\ell)} n^c$ is in fact polynomial in $n$.

**Theorem 4.** LONG CYCLE PARAMETERIZED BY MAX LEAF NUMBER *admits a polynomial kernel.*

*Proof.* Given an instance $(G, k, \ell)$ of LCML, we first check that $k$ does not exceed the number of vertices and that there are at most $4\ell$ vertices of degree at least three, or else return NO. If $G$ has more than $2^{\mathcal{O}(\ell)}$ vertices (using the concrete bound resulting from an implementation of Lemma 3), then we solve the instance in time $2^{\mathcal{O}(\ell)} \cdot n^c = \mathcal{O}(n^{c+1})$, and answer YES or NO accordingly. Otherwise let $B$ denote the set of vertices of degree at least three. If there are more than $\ell$ disjoint paths connecting any two vertices of $B$, then the max leaf number of $G$ exceeds $\ell$, and we return NO.

We replace each path connecting two vertices $b, b' \in B$, with internal vertices from $V(G) \setminus B$, by a single edge with an integer label denoting the number of internal vertices of the replaced path. We obtain a multigraph $G'$ in which some edges have an integer label. It is easy to see that cycles in $G$ correspond to cycles in $G'$ of the same length, when taking the integer labels into account (i.e. labeled edges are simply worth as much as that many internal vertices). Clearly, each label can be encoded in binary by at most $\log 2^{\mathcal{O}(\ell)} = \mathcal{O}(\ell)$ bits. Furthermore, we delete all paths that start in a vertex of $B$, have internal vertices from $V(G) \setminus B$, and end in a vertex of degree one; clearly those cannot be used by cycles in $G$.

We obtain a multigraph $G'$ with at most $4\ell$ vertices in $B$ and with at most $\ell$ edges between any two $B$-vertices. Thus we have $\mathcal{O}(\ell)$ vertices and $\mathcal{O}(\ell^3)$ integer labels of size $\mathcal{O}(\ell)$, for a total size of $\mathcal{O}(\ell^4)$ (this could be easily tightened, but it would not affect the result); clearly $k$ can also be encoded in $\mathcal{O}(\ell)$ bits.

We obtain an equivalent instance of a slightly different LONG CYCLE problem on multigraphs in which some edges may be labeled, but which is in NP. By the implied Karp reduction to LCML we obtain the claimed polynomial kernel (cf. [7]). Deviating from Bodlaender et al. [7] we do not use the versions with parameter encoded in unary, but observe the following: All instances of LONG CYCLE with $k$ exceeding the number of vertices are trivially NO and may be replaced by smaller dummy NO-instances, so the parameter value of the remaining instances is indeed polynomial in $\ell$ (as is the instance size, due to the Karp reduction). □

**Further problems parameterized by max leaf number.** A polynomial kernel for HAMILTONIAN CYCLE was already found by Fellows et al. [20]. Kernels for HAMILTONIAN PATH as well as DISJOINT CYCLES can be obtained in a similar way, by observing that the paths of degree-2 vertices can be reduced to having only one internal vertex. For LONG PATH it is again necessary to use the binary encoding trick for the path lengths.

**Corollary 2.** LONG PATH, HAMILTONIAN PATH, *and* DISJOINT CYCLES *parameterized by max leaf number admit polynomial kernels.*

For DISJOINT PATHS, i.e., finding $k$ disjoint paths connecting $k$ terminal pairs $(s_1, t_1), \ldots, (s_k, t_k)$, some more work is necessary on the paths between $B$ vertices, and on paths between $B$ vertices and the at most $\ell$ leaves.

**Theorem 5.** DISJOINT PATHS *parameterized by max leaf number admits a polynomial kernel.*

*Proof.* Let $(G, k, \{(s_1, t_1), \ldots, (s_k, t_k)\}, \ell)$ be an instance of DISJOINT PATHS PARAMETERIZED BY MAX LEAF NUMBER (DPML). Let $L$ be the set of leaves of $G$ and let $B$ contain all vertices of degree at least three. If $|L| > \ell$ or if $|B| > 4\ell$ we return NO as, by Lemma 2, the max leaf number of $G$ must exceed $k$. Similarly, we return NO if any vertex of $B$ has degree greater than $\ell$.

We will now reduce the number of terminals on paths with internal vertices from $O := V(G) \setminus (L \cup B)$, i.e., the set of vertices with degree exactly two. Let $P$ be such a path and assume that it contains at least five terminals. If follows that any terminal except the first and the last one on the path must reach its partner (e.g., $s_i$ must reach $t_i$) on the path $P$. This can be tested in polynomial time, and the instance be rejected if it is impossible. Afterwards the subpath spanned by those terminals can be deleted and be replaced by a subpath consisting just of the vertices of one new terminal pair, say $(s', t')$. By the same argumentation $s'$ and $t'$ must be connected on the path, hence the path cannot be used for other terminals. To see that the max leaf number does not increase, observe that the above modification is basically a contraction of some edges (terminals have no influence), and that it does not increase the number of connected components.

Otherwise, we replace all non-terminal vertices in $O$ by deleting them and adding an edge between their neighbors; it is clear that any path must pass through them, so this is equivalent. Afterwards, each path with internal vertices from $O$ has at most four internal vertices, allowing us to bound the total number of vertices by $|L| + |B| + 4\ell|B| \in \mathcal{O}(\ell^2)$ (the latter term accounts for the at most $\ell$ paths starting in any vertex of $B$). $\square$

## 4.3 Parameterization by a modulator to cluster graphs

In this section, we consider path and cycle problems parameterized by vertex-deletion distance from cluster graphs. To this end, alongside the input graph (and possibly further inputs) a modulator $X$ is provided such that $G - X$ is a cluster graph. Again we consider the LONG CYCLE problem.

> LONG CYCLE PARAMETERIZED BY A MODULATOR TO CLUSTER GRAPHS
> **Input:** A graph $G$, an integer $k$, and a set $X \subseteq V(G)$ such that $G - X$ is a disjoint union of cliques.
> **Parameter:** $\ell := |X|$.
> **Question:** Does $G$ contain a cycle of length at least $k$?

**Observation 2.** Cycles in $G$ contain vertices of at most $|X|$ cliques of $G - X$.

If a cycle uses at least one edge between vertices of a clique, then it can be easily extended to include all so far unused vertices of the clique. Hence, it can be seen that there is a preference for including the largest possible cliques and as many cliques as possible. This is used in the following reduction rule.

**Reduction Rule 2.** Given $(G, k, X)$, if there is a clique with at least $k$ vertices, or a vertex in $X$ with two neighbors in a clique of size $k-1$ then delete all other cliques and return the obtained instance $(G', k, X)$ (which is trivially YES).

Otherwise, for each vertex pair $\{u, v\}$ from $X$, mark $\ell + 1$ cliques that contain a shared neighbor of $u$ and $v$. Additionally, mark the $\ell + 1$ largest cliques containing vertices $p$ and $q$ with $u$ adjacent to $p$ and $v$ adjacent to $q$. Delete all unmarked cliques, obtaining a graph $G'$, and return $(G', k, X)$.

Clearly $G' - X$ is still a cluster graph, and if $G'$ contains a cycle of length at least $k$, then that cycle can also be found in its supergraph $G$. The following lemma completes the proof for safeness of Rule 2.

**Lemma 4.** *Let $(G', k, X)$ be obtained from an application of Rule 2 on $(G, k, X)$. If $G$ has a cycle of length at least $k$, then $G'$ has a cycle of length at least $k$.*

*Proof.* Assume that $G$ contains a cycle of length at least $k$. If the first part of Rule 2 applies, then $G'$ trivially contains a cycle of length $k$ and we are done. Otherwise, any cycle of length at least $k$ must contain at least two vertices of $X$.

Let $C$ be a cycle of length at least $k$ in $G$ which contains a minimum number of subpaths that are not contained in $G' - X$. Assume that $C$ contains vertices from at least one clique which is not in $G' - X$ (i.e. which was not marked) and let $K$ be such a clique of $G - X$. Let $P = (p_1, \ldots, p_r)$ denote one of the subpaths obtained by intersecting $C$ with the vertex set of $K$. Further, let $P_{uv}$ denote the extended subpath obtained by adding the vertices $u, v \in X$ which are adjacent to $P$ on $C$, w.l.o.g., $P_{uv} = (u, p_1, \ldots, p_r, v)$. Note that $u \neq v$, else $C = (u, p_1, \ldots, p_r)$ and the first case of Rule 2 would have applied.

If $r = 1$ and $P_{uv} = (u, p_1, v)$, then by the marking process of Rule 2 and as $K$ was not marked, there must be $\ell + 1$ cliques which contain a shared neighbor of $u$ and $v$ (which were marked and) which are present in $G'$. Hence, by Observation 2, at least one of those cliques, $K'$ say, was not used by $C$, and we may replace $p_1$ by any shared neighbor of $u$ and $v$ in $K'$. We obtain a cycle $C'$ that has one fewer subpath which is not in $G' - X$ contradicting our choice of $C$.

Now, if $r > 1$, then $P_{uv} = (u, p_1, \ldots, p_r, v)$ with $p_1 \neq p_r$. Hence, $K$ contains a neighbor of $u$ and a different neighbor of $v$. As $K$ is not marked, the reduction rule must have marked $\ell + 1$ other cliques which are at least as large as $K$, and which contain such neighbors. Again, by Observation 2, there must be a clique $K'$ among them which is not used by $C$. The clique $K'$ is at least as large as $K$ and it contains vertices $p$ and $q$ with $p$ adjacent to $u$ and $q$ adjacent to $v$. Hence, we may replace the subpath $P_{uv}$ of $C$ by a path from $u$ to $p$, followed by a path from $p$ to $q$ using all vertices of $K'$, and back to $v$; we call this $P'_{uv}$. Clearly, $P'_{uv}$ is at least as long as $P_{uv}$, since $P_{uv}$ could at most use all vertices of $K$ (there might be other subpaths of $C$ using $K$) and $K'$ has at least that many vertices. Thus, replacing $P_{uv}$ in $C$ by $P'_{uv}$ we obtain a cycle $C'$ of at least the same length, which uses one fewer subpath that is not contained in $G' - X$, contradicting the choice of $C$.

It follows that $C$ contains only vertices of $X$ and of marked cliques. Hence, it exists also in $G'$, completing the proof. □

For the remaining reduction we proceed similarly to Section 4.2: First, we show that we can without harm restrict to allowing only a small number of vertices in each clique for connecting to $X$, the remaining vertices are only needed to possibly extend the length of the cycle (i.e., they are visited after entering the clique and before leaving it). This allows for a straightforward FPT-algorithm of runtime $\mathcal{O}(\ell^{10\ell} \cdot n^c)$. Consequently, we may assume the number of vertices to be bounded by $\mathcal{O}(\ell^{10\ell})$, else solving the whole instance in time $\mathcal{O}(n^{c+1})$. Thus, the number of additional vertices in each clique may be encoded in binary, with coding length $\mathcal{O}(\ell \log \ell)$, giving a polynomial sized equivalent instance of a special version of our problem. Finally, we use a Karp reduction from this special version back to the basic problem, to obtain a polynomial kernel.

**Lemma 5.** *Given an instance $(G, k, X)$ one can in polynomial time identify $(\ell + 1)^3$ vertices in each clique of $G - X$, such that if $(G, k, X)$ is YES, then $G$ contains a cycle of length at least $k$ that enters and leaves cliques of $G - X$ only through the identified vertices.*

*Proof.* Given $(G, k, X)$ mark for each pair of vertices $u, v \in X$ up to $2\ell + 1$ shared neighbors in each clique. Furthermore, we mark for each vertex $v \in X$ up to $2\ell + 1$ neighbors in each clique.

Now, let us assume that $G$ contains a cycle of length at least $k$, and let $C$ be a cycle of length at least $k$ that enters and leaves cliques of $G - X$ through a minimum number of unmarked vertices (equivalently, $C$ uses a minimum number of edges between the modulator and unmarked vertices).

Assume that $C = (\ldots, u, p_1, \ldots, p_r, v, \ldots)$ where $u, v \in X$, all vertices $p_i$ are from the same clique $K$ of $G - X$, and at least $p_1$ is unmarked.

If $r = 1$, then $C = (\ldots, u, p_1, v, \ldots)$. As $p_1$ is unmarked, the clique $K$ must contain $2\ell + 1$ marked vertices that are shared neighbors of $u$ and $v$, say $q_1, \ldots, q_{2\ell+1}$. If at least one of them is not used by $C$, say $q_i$, then we may replace $p_1$ by $q_i$, and thereby obtain a cycle of the same length as $C$ that uses one less unmarked vertex to enter and leave cliques; a contradiction to the choice of $C$. So assume that $C$ uses all vertices $q_1, \ldots, q_{2\ell+1}$. By $\ell = |X|$ we know that at least one vertex, say $q_j$, is not adjacent to vertices of $X$ on $C$, but must be adjacent to other vertices of $K$. Therefore, we may swap the positions of $p_1$ and $q_j$ in $C$: Indeed, $q_j$ is adjacent to $u$ and $v$, and $p_1$ is adjacent to all other vertices of $K$ (and clearly it is not adjacent to $q_j$ on $C$). We obtain a cycle of the same length, which uses one less unmarked vertex to enter or leave cliques of $G - X$, contradicting our choice of $C$: The new cycle now enters a clique at $q_j$ instead of $p_1$.

Now, we consider the case that $r > 1$, so $C = (\ldots, u, p_1, \ldots, p_r, v, \ldots)$ with $p_1 \neq p_r$. Since $p_1$ is unmarked, there must be $2\ell + 1$ marked vertices in $K$ which are adjacent to $u$, say $q_1, \ldots, q_{2\ell+1}$. It follows that at least one of those vertices, say $q_j$, is not adjacent to $X$ on $C$. We can now apply the same switching arguments as in the previous paragraph to obtain a cycle $C'$ of the same length, which use one less unmarked vertex to enter or leave cliques of $G - X$, contradicting our choice of $C$.

It follows that $C$ enters and leaves cliques only through marked vertices. The marked vertices constitute the sets of vertices in the cliques as claimed by the lemma. There are at most $\ell \cdot (2\ell + 1) + \binom{\ell}{2} \cdot (2\ell + 1) \leq (\ell + 1)^3$ marked vertices per clique of $G - X$, as claimed. □

**Lemma 6.** *Given $(G, k, X)$, with $\ell := |X|$, a cycle of length at least $k$ in $G$ can be found in time $\mathcal{O}(\ell^{10\ell} \cdot n^c) = 2^{\mathcal{O}(\ell \log \ell)} n^{\mathcal{O}(1)}$ if it exists.*

*Proof.* We use Reduction Rule 2 to reduce the number of cliques in $G - X$ to $\mathcal{O}(\ell^3)$. Then we mark vertices according to Lemma 5. Assuming that $(G, k, X)$ is a YES-instance, it follows that there must be a cycle of length at least $k$ in $G$ which enters and leaves cliques only in marked vertices. The following brute force algorithm finds such a cycle if it exists:

1. If any clique has size at least $k$ then return YES. Else at least one vertex of $X$ must be used.

2. Try all ordered subsets $X' = \{x_1, \ldots, x_t\}$ of $X$ for the intersection of $C$ with $X$.

3. For all pairs $(x_i, x_{i+1})$ (including $(x_t, x_1)$) try all cliques that contain neighbors of $x_i$ and $x_{i+1}$, or, if possible, try the edge $\{x_i, x_{i+1}\}$ to connect $x_i$ and $x_{i+1}$ on $C$.

4. If a clique $K$ has been chosen to connect $x_i$ and $x_{i+1}$ on $C$, then try all marked vertices of the clique for entering and leaving $K$ on $C$, i.e., to find either vertices $p$ and $q$ with $C = (\ldots, x_i, p, \ldots, q, x_{i+1})$, or a single vertex $p$ for $C = (\ldots, x_i, p, x_{i+1}, \ldots)$ (and check adjacency of $p$ and possibly $q$ to $x_i$ and $x_{i+1}$).

5. Greedily connect the chosen vertices inside the cliques: For each clique where we have chosen the option $C = (\ldots, x_i, p, \ldots, q, x_{i+1}, \ldots)$ we can include all remaining vertices of the clique into our cycle. If only shared neighbors were chosen, then this is not possible.

It is easy to see that given the existence of any cycle $C'$ of length at least $k$ which enters and leaves cliques only through marked vertices, our simple algorithm must find a cycle of at least the same length: It will eventually try the same choice and order of vertices from $X$, and pick the same marked vertices to connect them to cliques. At that point, it is easy to see, that the greedy connection of the chosen marked vertices must give a cycle of at least the same length. Clearly, if our algorithm finds a cycle of length at least $k$, then the instance is YES.

We close by giving an upper bound on the dependence of $\ell$ in the runtime. It is easy to see that the dependence on the input size is (a low degree) polynomial.

1. This step can be performed in polynomial time.

2. There are less than $(\ell+1)^\ell = \mathcal{O}(\ell^\ell)$ ordered subsets $X'$ of $X$.

3. We pick up to $\ell$ cliques (possibly with repetition) out of the $\mathcal{O}(\ell^3)$ cliques of $G - X$, for a total of $\mathcal{O}(\ell^{3\ell})$ choices.

4. Up to $\ell$ times, we pick two marked vertices among the $(\ell+1)^3$ vertices, i.e., we have $\mathcal{O}(\ell^{6\ell})$ choices.

5. This step can be performed in time polynomial in the input size.

This gives a total runtime of $\mathcal{O}(\ell^{10\ell} \cdot n^c) = 2^{\mathcal{O}(\ell \log \ell)} n^{\mathcal{O}(1)}$. □

**Theorem 6.** LONG CYCLE PARAMETERIZED BY A MODULATOR TO CLUSTER GRAPHS *admits a polynomial kernel.*

*Proof.* Let $(G, k, X)$ be an input instance. W.l.o.g. we let the instance be reduced according to Reduction Rule 2, i.e., $G$ has at most $\mathcal{O}(\ell^3)$ cliques. If $G$ has at least $2^{\mathcal{O}(\ell \log \ell)}$ vertices, then the algorithm of Lemma 6 can be used to solve the instance in time $n \cdot n^{\mathcal{O}(1)} = n^{\mathcal{O}(1)}$. Otherwise, using $n < 2^{\mathcal{O}(\ell \log \ell)}$ we encode our instance in the following way, as an instance of a different problem:

- We mark vertices according to Lemma 5.

- In each clique, we replace all unmarked vertices by a single unmarked vertex with an integer label stating the number of unmarked vertices. In binary encoding, that label needs at most $\log n = \log(2^{\mathcal{O}(\ell \log \ell)}) = \mathcal{O}(\ell \log \ell)$ bits.

For the so-encoded instance we ask for a path of length at least $k$ which enters and leaves cliques only through marked vertices, but we account the labeled vertices as subpaths with a number of vertices equal to their label. Clearly, this is equivalent to the original question.

In addition to the vertices of $X$, the instance has at most $\mathcal{O}(\ell^3)$ cliques each with at most $(\ell+1)^3 + 1$ vertices plus one integer label of bit size at most $\mathcal{O}(\ell \log \ell)$ per clique. Clearly, this gives a total size which is polynomial in $\ell$.

Finally, we observe that the question which we ask about this instance can be answered in NP. As LONG CYCLE is NP-complete, it remains so even when we append the distance to a cluster graph in unary as well as a reasonable encoding of a modulator $X$ as both are bounded by $n$. Thus there is a Karp reduction from our alternative problem to LONG CYCLE PARAMETERIZED BY A MODULATOR TO CLUSTER GRAPHS. By standard arguments (e.g. see [7]) this implies that the latter problem admits a polynomial kernel. □

**Further problems parameterized by a modulator to cluster graphs.** For LONG PATH, HAMILTONIAN CYCLE, and HAMILTONIAN PATH polynomial kernels follow directly from Theorem 6 via straightforward reductions to the LONG CYCLE problem. However, it can be easily observed that for HAMILTONIAN CYCLE and HAMILTONIAN PATH the encoding trick is not necessary. Indeed all the unmarked vertices can always be assumed to occur in a single subpath of the final Hamiltonian cycle or path. Hence, it suffices to keep only a single unmarked vertex per clique. This saves the argument via Karp reductions. In fact, the marking argument can be tightened too, by using the matching approach from Section 4.1.

**Corollary 3.** LONG PATH, HAMILTONIAN CYCLE, *and* HAMILTONIAN PATH *parameterized by a modulator to cluster graphs admit polynomial kernels.*

*Proof.* For an instance $(G, k, X)$ of LONG PATH parameterized by a modulator from cluster graphs, simply add a universal vertex $v$ adjacent to all vertices of $G$ and ask for a cycle of length $k + 1$. Furthermore, the vertex $v$ is added to the modulator $X$, increasing the parameter value by one. Theorem 6 now creates an intermediate instance of size $\mathcal{O}(|X|^6)$ which reduces back to LONG PATH by a Karp reduction. (Note that we may also remove the universal vertex to get an instance of LONG PATH with labeled vertices. Of course this is a compression, or generalized kernelization.)

It is straightforward to get similar reductions for HAMILTONIAN CYCLE and HAMILTONIAN PATH by asking for long cycles or path respectively of length (at least) equal to the total number of vertices. However, it can be easily seen that following the proof of Theorem 6 for an obtained instance of LONG CYCLE, one may replace each labeled vertex by a single unlabeled one and update the target length (such that it always matches the number of vertices). Indeed, any cycle using all vertices can be modified to contain all unmarked vertices as a single subpath. The number of those vertices (i.e. the single label in each clique) is then immaterial. Thus one obtains graphs with $\mathcal{O}(|X|^6)$ vertices. It is straightforward to make the modifications to get back to instances of HAMILTONIAN CYCLE and HAMILTONIAN PATH respectively. □

We now turn our attention to DISJOINT PATHS and DISJOINT CYCLES. In both problems the length of paths or cycles is not important, but there maybe large numbers of paths and cycles inside each clique.

**Observation 3.** In the DISJOINT PATHS problem parameterized by a modulator $X$ to a cluster graph we may assume that all requested pairs lie entirely in $X$: First, it is clear that request pairs $(s_i, t_i)$ contained in the same clique of $G - X$, have the uniquely optimal path consisting only of $s_i$ and $t_i$. Thus, all such vertex pairs may be deleted (both from the graph and the list of requests). To satisfy any other request pair the corresponding path needs to intersect $X$, bounding the maximum feasible number of requests by $|X|$; else we reject the instance. Thus, including the vertices of all requests at most triples the size of $X$.

From this observation it is clear that a polynomial kernelization for DISJOINT PATHS can be achieved by marking (or matching) enough vertices to allow for the necessary paths. The case for DISJOINT CYCLES is similar: There are at most $|X|$ cycles which include vertices of $X$. All further cycles can be chosen as triangles inside single cliques. Again a marking procedure suffices, adding an additional step of removing triples of unmarked vertices from any single clique and each time reducing the target number of cycles by one (accounting for those trivial cycles not intersecting $X$).

**Corollary 4.** DISJOINT PATHS *and* DISJOINT CYCLES *parameterized by a modulator to a cluster graph admit polynomial kernels.*

# 5 Lower bounds for path and cycle problems

In this section we present kernelization lower bounds for the directed- and undirected variants of HAMILTONIAN CYCLE with structural parameters. The parameterizations we use are at least as large as the treewidth of the input graphs (or the underlying undirected graph, in the directed case) which shows that the parameterized problems for which we prove a kernel lower bound are indeed contained in FPT. We first develop a lower bound for DIRECTED HAMILTONIAN CYCLE PARAMETERIZED BY A MODULATOR TO BI-PATHS, and the show how to alter this construction to give lower bound for HAMILTONIAN CYCLE PARAMETERIZED BY A MODULATOR TO OUTERPLANAR GRAPHS. Our proofs use the technique of cross-composition [6], in which a kernel lower bound is obtained by showing that the logical OR of a series of instances of an NP-hard problem, can be embedded in a single instance of the parameterized target problem at a small parameter cost. We therefore start each subsection by defining an appropriate NP-hard problem to compose, and then give a cross-composition algorithm.

## 5.1 Directed Hamiltonian Cycle with a modulator to bi-paths

We start by defining the NP-hard problem which we will use in the cross-composition.

> HAMILTONIAN $s - t$ PATH IN DIRECTED BIPARTITE GRAPHS
> **Input:** A bipartite digraph $D$ with color classes $A = \{a_1, \ldots, a_{n_A}\}$ and $B = \{b_1, \ldots, b_{n_B}\}$ with $n_B = n_A + 1$ such that $N_D^-(b_1) = \emptyset$ and $N_D^+(b_{n_B}) = \emptyset$.
> **Question:** Does $D$ contain a directed Hamiltonian path which starts in $b_1$ and ends in $b_{n_B}$?

It is not difficult to show that this problem is NP-complete.

**Proposition 1.** HAMILTONIAN $s - t$ PATH IN DIRECTED BIPARTITE GRAPHS *is NP-complete.*

*Proof.* Membership in NP is trivial. We prove hardness by a reduction from HAMILTONIAN $s - t$ PATH which is a classical NP-complete problem [22, GT39]. On input an undirected graph $G$ with distinguished vertices $s, t$ construct a graph $G'$ on vertex set $V(G) \times \{1, 2, 3, 4\}$ with edges

$$\{\{v_1, v_2\}, \{v_2, v_3\}, \{v_3, v_4\} \mid v \in V(G)\} \cup \{\{v_1, u_4\}, \{v_4, u_1\} \mid \{u, v\} \in E(G)\}.$$

It is easy to verify that $G$ has a Hamiltonian $s - t$ path if and only if $G'$ has a Hamiltonian $s_1 - t_4$ path. The graph $G'$ is bipartite. To obtain an instance of HAMILTONIAN $s - t$ PATH IN DIRECTED BIPARTITE GRAPHS we build a digraph $D'$ by replacing the edges in $G'$ with bidirectional arcs, adding a new starting vertex $s^*$ with unique out-neighbor $s_1$, adding a vertex $w$ with in-neighbor $t_4$ and a new ending vertex $t^*$ with in-neighbor $w$. Taking the appropriate partite sets of the bipartition, labeling $s^*$ as $b_1$ and $t^*$ as $b_{n_B}$ we obtain an equivalent instance of HAMILTONIAN $s - t$ PATH IN DIRECTED BIPARTITE GRAPHS. □

Now we formally define the parameterized problem for which we will prove a kernel lower bound.

> DIRECTED HAMILTONIAN CYCLE PARAMETERIZED BY A MODULATOR TO BI-PATHS
> **Input:** A digraph $D$ and a modulator $X \subseteq V(D)$ such that $D - X \in$ BI-PATHS.
> **Parameter:** The size $|X|$ of the modulator.
> **Question:** Does $D$ have a directed Hamiltonian cycle?

The following propositions will be useful for verifying the correctness of the composition.

**Proposition 2.** *Let $D$ be a digraph with directed Hamiltonian cycle $C$. If $v \in V(D)$ is a vertex such that $N_D(v) = \{u, w\}$ then $C$ contains either the path $(u, v), (v, w)$ or $(w, v), (v, u)$.*

**Proposition 3.** *Let $D$ be a bipartite digraph with color classes $A = \{a_1, \ldots, a_{n_A}\}$ and $B = \{b_1, \ldots, b_{n_B}\}$ with $n_B = n_A + 1$. If $C \subseteq A(D)$ is a set of arcs such that:*

1. *$D[C]$ does not contain a directed cycle,*
2. *all vertices $a \in A$ are the head of one arc in $C$ and the tail of one arc in $C$,*
3. *no vertex $b \in B$ is the head of two arcs in $C$, or the tail of two arcs in $C$,*
4. *the vertex $b_1$ is the head of one arc in $C$ and the tail of no arc,*
5. *the vertex $b_{n_B}$ is the tail of one arc in $C$ and the head of no arc,*

*then $C$ is a Hamiltonian path from $b_1$ to $b_{n_B}$ in $D$.*

We can now give the cross-composition.

**Theorem 7.** DIRECTED HAMILTONIAN CYCLE PARAMETERIZED BY A MODULATOR TO BI-PATHS *does not admit a polynomial kernel unless* $NP \subseteq coNP/poly$.

*Proof.* By Theorem 1 and Proposition 1 it is sufficient to show that HAMILTONIAN $s - t$ PATH IN DIRECTED BIPARTITE GRAPHS cross-composes into DIRECTED HAMILTONIAN CYCLE PARAMETERIZED BY A MODULATOR TO BI-PATHS. We first give the polynomial equivalence relationship $\mathcal{R}$ to be used for the cross-composition. Fix some reasonable encoding of instances of HAMILTONIAN $s - t$ PATH IN DIRECTED BIPARTITE GRAPHS into an alphabet $\Sigma$ such that well-formed instances can be recognized in polynomial time. Now say that two strings in $\Sigma^*$ are equivalent under $\mathcal{R}$ if (a) they are both malformed instances, or (b) they encode instances $(D_1, A_1, B_1)$ and $(D_2, A_2, B_2)$ of HAMILTONIAN $s-t$ PATH IN DIRECTED BIPARTITE GRAPHS such that $|A_1| = |A_2|$ and $|B_1| = |B_2|$. It is not hard to verify that a set of strings which encode instances of up to $n$ vertices each, is partitioned into $\mathcal{O}(n)$ equivalence classes by $\mathcal{R}$, which is therefore a polynomial equivalence relationship.

It now suffices to give an algorithm which composes a sequence of instances of HAMILTONIAN $s - t$ PATH IN DIRECTED BIPARTITE GRAPHS which are equivalent under $\mathcal{R}$ into one instance of DIRECTED HAMILTONIAN CYCLE PARAMETERIZED BY A MODULATOR TO BI-PATHS. If the input consists of malformed instances then we can simply output a constant-size NO-instance. Hence in the remainder we may assume that the input contains $r$ well-formed instances $(D_1, A_1, B_1), \ldots, (D_r, A_r, B_r)$, and that $|A_i| = n_A$ and $|B_i| = n_B$ for $i \in [r]$ with $n_B = n_A + 1$. Label the vertices in each set $A_i$ as $a_{i,1}, \ldots, a_{i,n_A}$ and the vertices of a set $B_i$ as $b_{i,1}, \ldots, b_{i,n_B}$ for $i \in [r]$. Recall that instance $i$ asks whether $D_i$ has a Hamiltonian path from $b_{i,1}$ to $b_{i,n_B}$.

We construct a digraph $D^*$ as follows.

1. For $i \in [r]$, for $j \in [n_A]$ add vertices $a'_{i,j}, a''_{i,j}, a'''_{i,j}$ to $D^*$, and add arcs

$$(a'_{i,j}, a''_{i,j}), (a''_{i,j}, a'_{i,j}), (a''_{i,j}, a'''_{i,j}), (a'''_{i,j}, a''_{i,j}).$$

2. As the next step we add one-directional arcs to connect adjacent triples. For $i \in [r]$, for $j \in [n_A - 1]$ add the arc $(a'''_{i,j}, a'_{i,j+1})$.

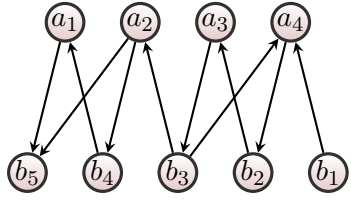

(a) Input instance $(D_1, A_1, B_1)$ of HAMILTONIAN $s-t$ PATH IN DIRECTED BIPARTITE GRAPHS.

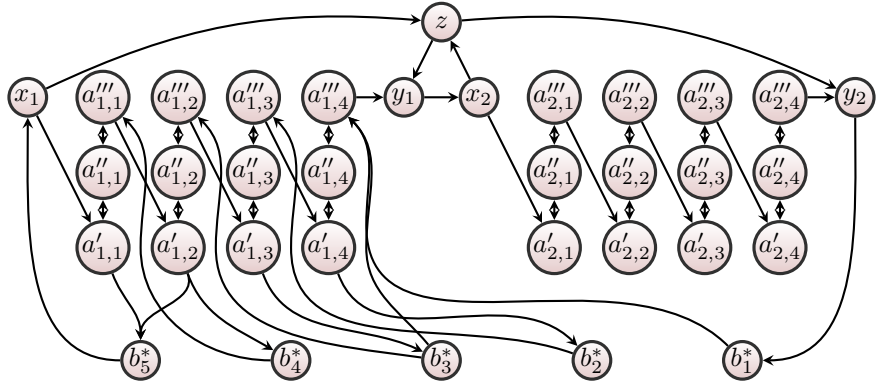

(b) Output instance of DIRECTED HAMILTONIAN CYCLE PARAMETERIZED BY A MODULATOR TO BI-PATHS.

Figure 1: An example of the lower-bound construction of Theorem 7 when composing $r = 2$ inputs with $n_A = 4$ and $n_B = 5$. (a) The first input instance. (b) Resulting output instance. The arcs between $\{b_1^*, \ldots, b_5^*\}$ and $\{a'_{2,j}, a''_{2,j}, a'''_{2,j} \mid j \in [4]\}$ which encode the second input $(D_2, A_2, B_2)$ have been omitted for readability.

3. For each instance $i \in [r]$ add two special vertices $x_i$ and $y_i$, together with arcs $(x_i, a'_{i,1})$ and $(a'''_{i,n_A}, y_i)$. For $i \in [r-1]$ add the arcs $(y_i, x_{i+1})$.

4. Observe that at this stage, $D^* \in \text{BI-PATHS}$. All vertices we add from this point on will go into the modulator $X^*$ such that $D^* - X^*$ will be a member of BI-PATHS.

5. We add a special vertex $z$ with arcs $(x_i, z)$ and $(z, y_i)$ for $i \in [r]$.

6. For $j \in [n_B]$ add a vertex $b_j^*$ to the graph $D^*$, and let $B^*$ be the set of these vertices. Add arcs $(y_r, b_1^*)$ and $(b_{n_B}^*, x_1)$.

7. As the last step of the construction we re-encode the behavior of the input graphs $D_i$ into the instance. For $i \in [r]$, for all arcs $(a_{i,j}, b_{i,h})$ in $A(D_i)$ add the arc $(a'_{i,j}, b_h^*)$ to $D^*$. For all arcs $(b_{i,j}, a_{i,h}) \in A(D_i)$ add $(b_j^*, a'''_{i,h})$ to $D^*$. This concludes the description of $D^*$, which is illustrated in Fig. 1.

Now define $X^* := \{z\} \cup B^*$. The output of the cross-composition is the instance $(D^*, X^*)$ of DIRECTED HAMILTONIAN CYCLE PARAMETERIZED BY A MODULATOR TO BI-PATHS. It is easy to verify that $D^* - X^* \in \text{BI-PATHS}$, and that the construction can be carried out in polynomial time. The parameter $|X^*|$ is bounded by $1 + n_B$ which is sufficiently small. It remains to prove that $D^*$ is YES if and only if one of the input instances is YES. Before proving this equivalence we establish some properties of $D^*$.

**Claim 1.** *Let $C \subseteq A(D^*)$ be a directed Hamiltonian cycle in $D^*$. The following must hold.*

1. *If $(a'''_{i,j}, a''_{i,j})$ is an arc on $C$ for some $i \in [r], j \in [n_A]$ then there are distinct indices $f, f'$ such that vertices $b^*_f, a'''_{i,j}, a''_{i,j}, a'_{i,j}, b^*_{f'}$ appear consecutively on $C$.*

2. *If $(x_i, a'_{i,1})$ is not an arc on $C$ for some $i \in [r]$, then none of the arcs $(a'''_{i,j}, a'_{i,j+1})$ for $j \in [n_A - 1]$ are contained in $C$, nor is the arc $(a'''_{i,n_A}, y_i)$.*

*Proof.* We prove the different components consecutively.

1. Assume that $(a'''_{i,j}, a''_{i,j})$ is an arc on $C$. By construction of $D^*$ the in-neighbors of $a'''_{i,j}$ which are not $a''_{i,j}$ are of the form $b^*_f$ for $f \in [n_B]$, and since $a''_{i,j}$ is used as the successor of $a'''_{i,j}$ this shows that the predecessor of $a'''_{i,j}$ must be some $b^*_f$. Since $N_{D^*}(a''_{i,j}) = \{a'_{i,j}, a'''_{i,j}\}$ we find by Proposition 2 that the vertices $a'''_{i,j}, a''_{i,j}, a'_{i,j}$ must be consecutive on $C$. Similarly as before, the construction of $D^*$ shows that the only out-neighbors of $a'_{i,j}$ different than $a''_{i,j}$ are of the form $b^*_{f'}$ for $f' \in [n_B]$. Since $a''_{i,j}$ is used as predecessor of $a'_{i,j}$ on $C$, it cannot be the successor and hence some $b^*_{f'}$ must be the successor of $a'_{i,j}$ on $C$ which shows that $b^*_f, a'''_{i,j}, a''_{i,j}, a'_{i,j}, b^*_{f'}$ are consecutive on $C$.

2. Assume that $(x_i, a'_{i,1})$ is not on $C$, but at least one of the arcs in the set $Z_i := \{(a'''_{i,j}, a'_{i,j+1}) \mid j \in [n_A - 1]\} \cup \{(a'''_{i,n_A}, y_i)\}$ is used on $C$. Let $j^*$ be smallest index such that the vertex $a'''_{i,j^*}$ is the head of an arc in $C \cap Z_i$. Then $a''_{i,j^*}$ is not the successor of $a'''_{i,j^*}$ on $C$ and by Proposition 2 it must therefore be its predecessor, showing that $a'_{i,j^*}, a''_{i,j^*}, a'''_{i,j^*}$ must be consecutive on $C$. If $j^* \geq 2$ then the only in-neighbors of $a'_{i,j^*}$ in $D^*$ are $\{a'''_{i,j^*-1}, a''_{i,j^*}\}$, and if $j^* = 1$ then the only in-neighbors are $\{x_i, a''_{i,j^*}\}$. By choice of $j^*$ as the *smallest* index from $Z_i$ which is the head of an arc in $C \cap Z_i$, the first in-neighbor of $a'_{i,j^*}$ cannot be its predecessor on $C$. But since $a''_{i,j^*}$ is its successor on $C$, that vertex cannot be its predecessor either. So $a'_{i,j^*}$ does not have a predecessor on $C$ which contradicts the assumption that $C$ is a Hamiltonian cycle, which concludes the proof of this part. □

We are now ready to prove that $(D^*, X^*)$ is YES if and only if one of the input instances is YES. For the first direction, assume that $D^*$ has a directed Hamiltonian cycle $C$. Since $N^-_{D^*}(z) = \{x_i \mid i \in [r]\}$ there is an index $i^*$ such that $x_{i^*}$ is the predecessor of $z$ on $C$, which shows that the arc $(x_{i^*}, a'_{i^*,1})$ is not used on $C$. By (2) this implies that there are no arcs $(a'''_{i^*,j}, a'_{i^*,j+1})$ for $j \in [n_A - 1]$ contained in $C$, nor is the arc $(a'''_{i^*,n_A}, y_{i^*})$. By construction of $D^*$ this implies that each vertex $a'''_{i^*,j}$ for $j \in [n_A]$ has $a''_{i^*,j}$ as its successor on $C$ (there is only one other option for the successor, which is explicitly excluded). Hence all the arcs $(a'''_{i^*,j}, a''_{i^*,j})$ are contained in $C$ for $j \in [n_A]$. By (1) this shows that for each $j \in [n_B]$ there are $b^*_f, b^*_{f'}$ such that $b^*_f, a'''_{i^*,j}, a''_{i^*,j}, a'_{i^*,j}, b^*_{f'}$ are consecutive on $C$. Now consider the set of arcs $C_{i^*}$ in $D_{i^*}$ defined as follows:

- If $b^*_f, a'''_{i^*,j}, a''_{i^*,j}, a'_{i^*,j}, b^*_{f'}$ are consecutive on $C$ then add the arcs $(b_{i^*,f}, a_{i^*,j})$ and $(a_{i^*,j}, b_{i^*,f'})$ to $C_{i^*}$.

Recall that by construction of $D^*$, the arc $(b^*_f, a'''_{i^*,j})$ is only present in $D^*$ when $(b_{i^*,f}, a_{i^*,j}) \in A(D_{i^*})$, and the arc $(a'_{i^*,j}, b^*_f)$ is only present in $D^*$ when $(a_{i^*,j}, b_{i^*,f}) \in A(D_{i^*})$, and hence all arcs added to $C_{i^*}$ by this definition are indeed present in $D_{i^*}$. We will show that the set $C_{i^*} \subseteq A(D_{i^*})$ satisfies all requirements of Proposition 3 for graph $D_{i^*}$, thereby showing that $D_{i^*}$ has a Hamiltonian path from $b_{i^*,1}$ to $b_{i^*,n_B}$.

If $C_{i^*}$ contains a directed cycle, then by construction of $C_{i^*}$ it follows that $C$ must contain a closed cycle on a vertex subset of $\{b_j^* \mid j \in [n_B]\} \cup \{a'_{i^*,j}, a''_{i^*,j}, a'''_{i^*,j} \mid j \in [n_A]\}$ showing that $C$ is not a Hamiltonian cycle in $D^*$; hence the set $C_{i^*}$ satisfies (1) of Proposition 3. The definition of $C_{i^*}$ directly shows that $C$ satisfies (2). If some vertex $b_{i^*,j}$ is the head or tail of two arcs in $C_{i^*}$ then the corresponding vertex $b_j^*$ is head or tail of two arcs in the Hamiltonian cycle $C$, which is not possible; hence (3) is satisfied. Since the definition of HAMILTONIAN $s - t$ PATH IN DIRECTED BIPARTITE GRAPHS guarantees that $b_{i^*,1}$ does not have in-arcs in $D_{i^*}$, and that $b_{i^*,n_B}$ has no out-arcs in $D_{i^*}$, the vertex $b_1^*$ cannot occur as successor to a vertex $a'_{i^*,j}$ and vertex $b_{n_B}^*$ cannot occur as predecessor of a vertex $a'''_{i^*,j}$. Therefore $C_{i^*}$ contains no arcs leading into $b_{i^*,1}$ and no arcs leading out of $b_{i^*,n_B}$, proving that the last condition is also satisfied. By the proposition this proves that $C_{i^*}$ is a Hamiltonian $b_{i^*,1} - b_{i^*,n_B}$ path in $D_{i^*}$ which proves this direction of the equivalence.

The other direction is straight-forward. Assume that $C_{i^*}$ is a Hamiltonian path from $b_{i^*,1}$ to $b_{i^*,n_B}$ in $D_{i^*}$. We construct a Hamiltonian cycle $C$ in $D^*$ as follows.

- For $i \neq i^*$ add all arcs between consecutive vertices of $x_i, a'_{i,1}, a''_{i,1}, a'''_{i,1}, a'_{i,2}, \ldots, a'''_{i,n_A-1}, a'_{i,n_A}, a''_{i,n_A}, a'''_{i,n_A}, y_i$ to $C$.
- Add all arcs $(a'''_{i^*,j}, a''_{i^*,j}), (a''_{i^*,j}, a'_{i^*,j})$ for $j \in [n_A]$ to $C$.
- Add all arcs $(y_i, x_{i+1})$ for $i \in [r-1]$ to $C$.
- Add the arcs $(b_{n_B}^*, x_1)$ and $(y_r, b_1^*)$ to $C$.
- For each arc $(b_{i^*,f}, a_{i^*,j}) \in C_{i^*}$ add $(b_f^*, a'''_{i^*,j})$ to $C$.
- For each arc $(a_{i^*,j}, b_{i^*,f}) \in C_{i^*}$ add $(a'_{i^*,j}, b_f^*)$ to $C$.

Using the construction of $D^*$ is it straight-forward to verify that $C$ is a Hamiltonian cycle in $D^*$, which proves the reverse direction of the equivalence and concludes the proof. □

It is not difficult to see that the proof of Theorem 7 can be adapted to give a kernel lower bound for the variant where we are looking for a Hamiltonian path instead of a Hamiltonian cycle; these bounds in turn imply that the versions where we are looking for a long path or cycle (instead of one which is Hamiltonian) are at least as hard to kernelize, as is the case when we want to find a long $s - t$ path or a long cycle through a given vertex.

## 5.2 Hamiltonian Cycle with a modulator to outerplanar graphs

For the cross-composition of this section we will use the following variant of HAMILTONIAN PATH on undirected bipartite graphs.

> HAMILTONIAN $s - t$ PATH IN BIPARTITE GRAPHS
> **Input:** A bipartite graph $G$ with color classes $A = \{a_1, \ldots, a_{n_A}\}$ and $B = \{b_1, \ldots, b_{n_B}\}$ with $n_B = n_A + 1$ such that $|N_G(b_1)| = |N_G(b_{n_B})| = 1$.
> **Question:** Does $G$ contain a Hamiltonian path which starts in $b_1$ and ends in $b_{n_B}$?

NP-completeness of this problem follows from the construction of Proposition 1.

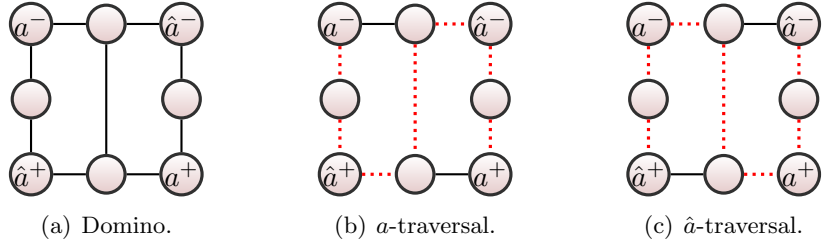

(a) Domino.     (b) $a$-traversal.     (c) $\hat{a}$-traversal.

Figure 2: (a) The domino gadget with four terminal vertices $a^-, a^+, \hat{a}^-, \hat{a}^+$. If graph $G$ contains the domino as a subgraph such that only these terminals are connected to the remainder of the graph, then any Hamiltonian cycle for $G$ must use the $a$-traversal (b) or $\hat{a}$-traversal (c) of the domino.

**Proposition 4.** *The* HAMILTONIAN $s - t$ PATH IN BIPARTITE GRAPHS *problem is NP-complete.*

The problem for which we prove a lower bound is formally defined as follows.

> HAMILTONIAN CYCLE PARAMETERIZED BY A MODULATOR TO OUTERPLANAR GRAPHS
> **Input:** A graph $G$ and a modulator $X \subseteq V(G)$ such that $G - X \in$ OUTERPLANAR.
> **Parameter:** The size $|X|$ of the modulator.
> **Question:** Does $G$ have a Hamiltonian cycle?

We can modify the construction of Theorem 7 to give a lower bound for HAMILTONIAN CYCLE PARAMETERIZED BY A MODULATOR TO OUTERPLANAR GRAPHS. Our main tool is the "domino" gadget of Fig. 2, which a Hamiltonian cycle must visit in one of two specific ways. This domino will be used to simulate directed edges.

**Proposition 5.** *Let $G$ be an undirected graph containing the domino as an induced subgraph, such that only the vertices labeled $a^+, a^-, \hat{a}^+, \hat{a}^-$ have neighbors outside the domino. Then any Hamiltonian cycle in $G$ must either (a) contain an a-traversal of the domino, which is path between $a^+$ and $a^-$, visiting all other vertices of the domino in between or (b) contain an $\hat{a}$-traversal of the domino, which is a path between $\hat{a}^+$ and $\hat{a}^-$, visiting all other vertices in between.*

**Theorem 8.** HAMILTONIAN CYCLE PARAMETERIZED BY A MODULATOR TO OUTERPLANAR GRAPHS *admits no polynomial kernel unless* $NP \subseteq coNP/poly$.

*Proof.* By Theorem 1 and Proposition 4 it is sufficient to show that HAMILTONIAN $s - t$ PATH IN BIPARTITE GRAPHS cross-composes into HAMILTONIAN CYCLE PARAMETERIZED BY A MODULATOR TO OUTERPLANAR GRAPHS. By arguments similar to that in the proof of Theorem 7 we can define a suitable polynomial equivalence relationship $\mathcal{R}$ such that it is sufficient to give an algorithm which, given $r$ well-formed instances $(G_1, A_1, B_1), \ldots, (G_r, A_r, B_r)$ of HAMILTONIAN $s - t$ PATH IN BIPARTITE GRAPHS such that $|A_i| = n_A$, $|B_i| = n_B$ with $n_B = n_A + 1$ for $i \in [r]$ outputs in polynomial time an instance $(G^*, X^*)$ of HAMILTONIAN CYCLE PARAMETERIZED BY A MODULATOR TO OUTERPLANAR GRAPHS which is YES if and only if one of the input instances is YES, and such that $|X^*|$ is polynomial in the size of the largest input instance plus $\log r$.

The construction is similar to that of the graph $D^*$ in Theorem 7, with the difference that we are now using undirected graphs and that we use an outerplanar gadget to simulate directions of arcs. Assume $A_i = \{a_{i,1}, \ldots, a_{i,n_A}\}$ and $B_i = \{b_{i,1}, \ldots, b_{i,n_B}\}$ for $i \in [r]$. We build a graph $G^*$ as follows.

1. For $i \in [r]$, for $j \in [n_A]$ add a copy $O_{i,j}$ of the domino graph (Fig. 2) to $G^*$ and label its terminals by $a_{i,j}^-, a_{i,j}^+, \hat{a}_{i,j}^-$ and $\hat{a}_{i,j}^+$.

2. As the next step we add edges to connect adjacent dominos. For $i \in [r]$, for $j \in [n_A - 1]$ add the edge $\{a_{i,j}^+, a_{i,j+1}^-\}$.

3. For each instance $i \in [r]$ add three special vertices $x_i, y_i$, and $w_i$, together with edges $\{w_i, x_i\}$, $\{x_i, a_{i,1}^-\}, \{a_{i,n_A}^+, y_i\}$. For $i \in [r-1]$ add the edges $\{y_i, w_{i+1}\}$.

4. Observe that at this stage, $G^* \in \text{OUTERPLANAR}$. All vertices we add from this point on will go into the modulator $X^*$ such that $G^* - X^*$ will be a member of OUTERPLANAR.

5. We add three special vertices $z^-, z, z^+$ with edges with $\{z^-, z\}$ and $\{z, z^+\}$. Furthermore, we add edges $\{x_i, z^-\}$ and $\{z^+, y_i\}$ for $i \in [r]$.

6. For $j \in [n_B]$ add a vertex $b_j^*$ to the graph $G^*$, and let $B^*$ be the set of these vertices. Add edges $\{y_r, b_1^*\}$ and $\{b_{n_B}^*, w_1\}$.

7. As the last step of the construction we re-encode the behavior of the input graphs $D_i$ into the instance. For $i \in [r]$, for each edge $\{a_{i,j}, b_{i,f}\}$ in $E(G_i)$ add the edges $\{\hat{a}_{i,j}^-, b_f^*\}$ and $\{\hat{a}_{i,j}^+, b_f^*\}$ to $G^*$. This concludes the description of $G^*$.

Now define $X^* := \{z^-, z, z^+\} \cup B^*$. The output of the cross-composition is the instance $(G^*, X^*)$ of HAMILTONIAN CYCLE PARAMETERIZED BY A MODULATOR TO OUTERPLANAR GRAPHS. It is easy to verify that $G^* - X^* \in \text{OUTERPLANAR}$, and that the construction can be carried out in polynomial time. The parameter $|X^*|$ is bounded by $1 + n_B$ which is sufficiently small. It remains to prove that $G^*$ is YES if and only if one of the input instances is YES. We first prove some relevant properties of $G^*$.

**Claim 2.** *Let $C \subseteq E(G^*)$ be a Hamiltonian cycle in $G^*$. The following must hold.*

1. *If $C$ uses an $\hat{a}$-traversal of some domino $O_{i,j}$ for $i \in [r]$ and $j \in [n_A]$ then there are distinct indices $f, f'$ such that $b_f$, the vertices of the domino, and $b_{f'}$ appear consecutively on $C$.*

2. *If $\{x_i, a_{i,1}^-\}$ is not an edge on $C$ for some $i \in [r]$, then none of the edges $\{a_{i,j}^+, a_{i,j+1}^-\}$ for $j \in [n_A - 1]$ are contained in $C$, nor is the edge $\{a_{i,n_A}^+, y_i\}$.*

3. *There is an index $i^* \in [r]$ such that $\{x_i, a_{i,1}^+\}$ is not used on $C$.*

*Proof.* We prove the different components consecutively.

1. Suppose $C$ uses an $\hat{a}$-traversal of some domino $O_{i,j}$. By construction of $G^*$ we know that the only neighbors of $\hat{a}_{i,j}^+$ and $\hat{a}_{i,j}^-$ which are not contained in the domino are of the form $b_f^*$ for some $f \in [n_B]$. Since all vertices of the domino are used on the cycle $C$ in an $\hat{a}$-traversal (see Fig. 2) and the vertices $\hat{a}_{i,j}^+$ and $\hat{a}_{i,j}^-$ each have only one neighbor on $C$ within the domino, each of these vertices must have a neighbor outside the domino on $C$ and hence these must be of the form $b_f^*$ and $b_{f'}^*$. We must have $f \neq f'$ otherwise we would have a closed cycle containing the domino $O_{i,j}$ and a single vertex $b_f^*$, and this cycle would not be Hamiltonian since it would not visit the vertex $z$.

2. Assume that $\{x_i, a_{i,1}^-\}$ is not on $C$, but at least one of the edges in the set $Z_i := \{\{a_{i,j}^+, a_{i,j+1}^-\} \mid j \in [n_A - 1]\} \cup \{\{a_{i,n_A}^+, y_i\}\}$ is used on $C$. Let $j^*$ be smallest index such that the vertex $a_{i,j^*}^+$ is the endpoint of an edge in $C \cap Z_i$. Since $a_{i,j^*}^+$ is endpoint of an edge in $C$ and the other endpoint of this edge does not lie in the domino $O_{i,j^*}$, it follows that $C$ contains at most one edge in the domino $O_{i,j^*}$ which is incident on $a_{i,j^*}^+$. This rules out the possibility that $C$ makes an $\hat{a}$-traversal of $O_{i,j^*}$, since that requires two edges within the domino incident on $a_{i,j^*}^+$. Because the construction of $G^*$ together with Proposition 5 ensures there are only two possible traversals of the domino, we know that $C$ must use an $a$-traversal of $O_{i,j^*}$. This traversal uses exactly one edge of the domino incident on $a_{i,j^*}^-$. Since a Hamiltonian cycle must contain exactly two edges incident on every vertex, this shows that $C$ must contain some edge incident on $a_{i,j^*}^-$ which is not in the domino $O_{i,j^*}$. But by construction of $G^*$ there is only one such edge: if $j^* = 1$ then this edge is $\{x_i, a_{i,1}^-\}$ and otherwise this edge is $\{a_{i,j^*-1}^+, a_{i,j^*}^-\}$. But by the assumption at the start of the proof and our choice of $j^*$, this edge is not contained in $C$. Hence there is only one edge incident on $a_{i,j^*}^-$ contained in $C$, contradicting the Hamiltonicity of $C$.

3. The Hamiltonian cycle $C$ must contain two edges incident on every vertex. By construction of $G^*$ the vertex $z^-$ is incident on the edge $\{z^-, z\}$ and on edges $\{x_i, z^-\}$ for $i \in [r]$. Hence if $C$ contains two edges incident on $z^-$ then at least one is of the form $\{x_{i^*}, z^-\}$ for $i^* \in [r]$. Now consider the vertex $w_{i^*}$, which has degree two by construction. Therefore both its incident edges must be contained in $C$, and in particular $\{w_{i^*}, x_{i^*}\}$ is contained in $C$. So $C$ contains two edges incident on $x_{i^*}$ which are unequal to the edge $\{x_{i^*}, a_{i^*,1}^-\}$ and since only two edges incident to each vertex are contained in $C$, it follows that $C$ does not contain $\{x_{i^*}, a_{i^*,1}^-\}$. □

Armed with these claims we prove that $G^*$ has a Hamiltonian cycle if and only if one of the input instances $G_{i^*}$ has a Hamiltonian path starting in $b_1$ and ending in $b_{n_B}$. First assume $G^*$ has a Hamiltonian cycle $C \subseteq E(G^*)$. By (3) there is an index $i^* \in [r]$ such that $\{x_i, a_{i^*,1}^-\} \notin C$. By (2) this implies there are no edges $\{a_{i^*,j}^+, a_{i^*,j+1}^-\}$ for $j \in [n_A - 1]$ contained in $C$, nor is the edge $\{a_{i^*,n_A}^+, y_i\}$. This shows that for each vertex $a_{i^*,j}^-$ the only edges incident on $a_{i^*,j}^-$ which can be contained in $C$, are those of the domino $O_{i^*,j}$ which implies by Proposition 5 that $C$ must do an $\hat{a}$-traversal of all dominos $O_{i^*,j}$ for $j \in [n_A]$. By (1) each traversal of a domino is preceded and succeeded by distinct vertices $b_f^*, b_{f'}^*$ and by construction of $G^*$ this can only occur if $\{a_{i^*,j}, b_{i^*,f}\}$ and $\{a_{i^*,j}, b_{i^*,f'}\}$ are edges of $G_{i^*}$. Since vertices $b_{i^*,1}$ and $b_{i^*,n_B}$ have degree one in $G_{i^*}$ by definition of HAMILTONIAN $s - t$ PATH IN BIPARTITE GRAPHS, and since $n_B = n_A + 1$, it now follows by reasoning similar to that of Theorem 7 that the set $C_{i^*} := \{\{a_{i^*,j}, b_{i^*,f}\} \mid b_f^* \text{ precedes or succeeds domino } O_{i^*,j} \text{ on } C\}$ is a Hamiltonian path in $G_{i^*}$ between vertices $b_{i^*,1}$ and $b_{i^*,n_B}$.

For the reverse direction, assume that $G_{i^*}$ has a Hamiltonian path $C_{i^*}$ starting in $b_{i^*,1}$ and ending in $b_{i^*,n_B}$. Construct a Hamiltonian cycle in $G^*$ as follows.

- For $i \neq i^*$ add the edges $\{w_i, x_i\}$, $\{x_i, a_{i,1}^-\}$, $\{a_{i,n_A}^+, y_i\}$, and $\{a_{i,j}^+, a_{i,j+1}^-\}$ for $j \in [n_A - 1]$ to $C$.

- For $i \neq i^*$ add the edges on the $a$-traversals of all dominos $O_{i,j}$ with $j \in [n_A]$ to $C$.

- Add the edges $\{w_{i^*}, x_{i^*}\}$, $\{x_{i^*}, z^-\}$, $\{z^-, z\}$, $\{z, z^+\}$, and $\{z^+, y_{i^*}\}$ to $C$. Also add all edges on $\hat{a}$-traversals of the dominos $O_{i^*,j}$ for $j \in [n_A]$ to $C$.

- Add all edges $\{y_i, x_{i+1}\}$ for $i \in [r - 1]$ to $C$.

- Add the edges $\{b^*_{n_B}, w_1\}$ and $\{y_r, b^*_1\}$ to $C$.

- For $j \in [n_A]$, if the edges of $G_{i^*}$ incident on $a_{i^*,j}$ in the Hamiltonian path $C_{i^*}$ are $\{a_{i^*,j}, b_{i^*,f}\}$ and $\{a_{i^*,j}, b_{i^*,f'}\}$ then add the edges $\{\hat{a}^-_{i^*,j}, b^*_f\}$ and $\{\hat{a}^+_{i^*,j}, b^*_{f'}\}$ to $C$. (The ordering of $b^*_f$ and $b^*_{f'}$ is immaterial since the domino can be traversed in either direction.)

Using the construction of $G^*$ is it straight-forward to verify that $C$ is a Hamiltonian cycle in $G^*$, which proves the reverse direction of the equivalence and concludes the proof. □

## 6 Finding paths with respect to forbidden pairs

In this section we study multiple parameterizations of several variants of path problems involving forbidden pairs. The first version we consider is defined as follows.

> $s - t$ PATH WITH FORBIDDEN PAIRS PARAMETERIZED BY A VERTEX COVER OF $G$
> **Input:** A graph $G$, distinct vertices $s, t \in V(G)$, a set $H \subseteq \binom{V(G)}{2}$ of forbidden pairs, and a vertex cover $X$ of $G$.
> **Parameter:** $\ell := |X|$.
> **Question:** Is there an $s - t$ path in $G$ which contains at most one vertex of each pair $\{u, v\} \in H$?

We can give evidence that this problem is not fixed-parameter tractable.

**Theorem 9.** $s - t$ PATH WITH FORBIDDEN PAIRS PARAMETERIZED BY A VERTEX COVER OF $G$ is hard for W[1].

*Proof.* We give a parameterized reduction from the W[1]-hard $k$-MULTICOLORED CLIQUE problem [19]. Let $(G, k, \phi)$ be an instance of $k$-MULTICOLORED CLIQUE, where $\phi : V(G) \to \{1, \ldots, k\}$ assigns each vertex of $G$ a color from 1 to $k$ and the task is to decide whether $G$ contains a clique with one vertex of each of the $k$ colors.

We construct a graph $G'$ as follows. We first make an independent set on the vertices $V(G)$, and add a forbidden pair for each pair of vertices from $V(G)$ that are either not adjacent in $G$ or which have the same color according to $\phi$. Then we add $k + 1$ vertices $v_0, v_1, \ldots, v_k$ which will form the vertex cover. We connect vertex $v_0$ to all vertices of the independent set that have color 1 according to $\phi$. For all $i \in \{1, \ldots, k - 1\}$, we connect $v_i$ to all vertices that have colors $i$ or $i + 1$. Finally, we connect $v_k$ to all vertices of color $k$. We set $s := v_0$ and $t := v_k$, and return the instance $(G', v_0, v_k, X' := \{v_0, \ldots, v_k\})$. We note that this can be easily performed in polynomial time, and that the parameter value, i.e., $k + 1$, is bounded by a function of $k$.

It is easy to see that every $s - t$ path has $2k + 1$ vertices, and uses all vertices $v_0, \ldots, v_k$ as well as one vertex of each color. Note that the colors are not part of the produced instance, but the forbidden pairs are used to encode this property. The latter vertices can also not be non-adjacent in $G$, and hence they correspond to a multicolored clique of size $k$. Similarly, given such a multicolored clique it is straightforward to find an $s - t$ path of length $2k + 1$ in $G'$ that respects the forbidden pairs.

Thus W[1]-hardness of $s - t$ PATH WITH FORBIDDEN PAIRS PARAMETERIZED BY A VERTEX COVER OF $G$ follows. □

Let us now consider some variations of the problem. The problems SHORTEST $s-t$ PATH WITH FORBIDDEN PAIRS and LONGEST $s-t$ PATH WITH FORBIDDEN PAIRS are defined similarly as $s-t$ PATH WITH FORBIDDEN PAIRS, with the difference that there is an extra integer $k$ in the input and we are asking for an $s-t$ path containing at most or at least $k$ vertices. In LONGEST PATH WITH FORBIDDEN PAIRS we omit the inputs $s$ and $t$, and are looking for *any* sufficiently long path, regardless of its endpoints. The related problem SHORTEST PATH WITH FORBIDDEN PAIRS is not interesting, since its solution always consist of a path containing a single vertex.

It can be easily verified that asking for a path on at least $2k+1$ vertices, regardless of its endpoints, in the construction for the proof of Theorem 9 leads also to a path whose vertices from the independent set made of $V(G)$ correspond to a multicolored clique in $G$. Also, since asking for a long or short $s-t$ path is as least as hard as asking for the existence of an $s-t$ path, we obtain the following results as a corollary.

**Corollary 5.** *The problems* SHORTEST $s-t$ PATH WITH FORBIDDEN PAIRS, LONGEST $s-t$ PATH WITH FORBIDDEN PAIRS *and* LONGEST PATH WITH FORBIDDEN PAIRS, *are hard for W[1] when parameterized by a vertex cover of $G$.*

Clearly, the hardness of the path problems with forbidden pairs stems from the extra structure of the forbidden pairs $H$, which is not taken into account when considering structural parameters of $G$. In the following we consider the effect of parameterizing by the structure of the graph $G \cup H$ (i.e., $G$ with an added edge for every forbidden pair).

Using the optimization version of Courcelle's Theorem applied to *structures* of bounded treewidth [2, 4, 8, 12, 13] (instead of graphs, as is more common) it is not difficult to obtain an FPT result parameterized by the treewidth of $G \cup H$. If we take a structure on the base set $V(G) \cup E(G) \cup H$ which encodes an instance through unary predicates which say whether an object is a vertex of $G$, edge of $G$, or forbidden pair in $H$, and through binary predicates which give the incidence between vertices and edges or pairs, then the treewidth of the structure equals the treewidth of $G \cup H$. For such a representation it is well-known how to devise an MSOL formula which asks whether there exists a set of edges which forms a path between $s$ and $t$, and such that no two vertices incident on such an edge form a forbidden pair. Using standard extensions of MSOL we may also maximize or minimize the size of a set of edges which forms an $s-t$ path respecting forbidden pairs, obtaining the following result.

**Proposition 6.** *The problems* SHORTEST $s-t$ PATH WITH FORBIDDEN PAIRS, LONGEST $s-t$ PATH WITH FORBIDDEN PAIRS, *and* LONGEST PATH WITH FORBIDDEN PAIRS, *are fixed-parameter tractable parameterized by the treewidth of $G \cup H$.*

For the case of SHORTEST $s-t$ PATH WITH FORBIDDEN PAIRS the structure of $G$ is actually not so important for the complexity of the problem: it is sufficient to parameterize by a vertex cover of the graph on the edge set $H$ to obtain fixed-parameter tractability, as demonstrated by the following theorem.

**Theorem 10.** SHORTEST $s-t$ PATH WITH FORBIDDEN PAIRS PARAMETERIZED BY A VERTEX COVER OF $H$ *is fixed-parameter tractable.*

*Proof.* Given a graph $G$ with endpoints $s, t$ and forbidden pairs $H$ such that $X$ is a vertex cover of $H$, we describe how to find the shortest $s-t$ path which avoids forbidden pairs. For all subsets $X' \subseteq X$ which do not contain a forbidden pair, we compute the shortest $s-t$ path which intersects $X$ only in $X'$ as follows. Let $Y$ be the vertices which form a forbidden pair with a member of $X'$: then the desired path is a shortest $s - t$ path in the graph $G - (X \setminus X') - Y$. Since a shortest path in this graph can be found in linear time using breadth-first search, we solve the problem in $\mathcal{O}^*(2^{|X|})$ time by returning the shortest $s - t$ path found over all choices of $X' \subseteq X$. □

It is easy to see that the positive news of Theorem 10, membership in FPT parameterized by a vertex cover of $H$, cannot be extended to LONGEST $s - t$ PATH WITH FORBIDDEN PAIRS since the latter problem is already NP-complete when there are no forbidden pairs. We mention without proof that $s - t$ PATH WITH FORBIDDEN PAIRS is NP-complete when the graph induced by $H$ is a matching, showing that we cannot improve the parameterization by a vertex cover of $H$ to the treewidth of $H$.

As the final topic of this section we will consider the kernelization complexity of forbidden path problems, obtaining a super-polynomial lower bound on the kernel size when parameterizing by a vertex cover of $G \cup H$.

**Theorem 11.** $s - t$ PATH WITH FORBIDDEN PAIRS PARAMETERIZED BY A VERTEX COVER OF $G \cup H$ admits no polynomial kernel unless $NP \subseteq coNP/poly$.

*Proof.* We consider the classical problem $s - t$ PATH WITH FORBIDDEN PAIRS where we simply drop the set $X$ from the problem description. We show that $s - t$ PATH WITH FORBIDDEN PAIRS cross-composes into $s - t$ PATH WITH FORBIDDEN PAIRS PARAMETERIZED BY A VERTEX COVER OF $G \cup H$, which suffices to prove the claim by Theorem 1 since the construction of Theorem 9 shows that $s - t$ PATH WITH FORBIDDEN PAIRS is NP-complete.

Let $\mathcal{R}$ be a polynomial equivalence relation under which two well-formed instances $(G_1, s_1, t_1, H_1)$ and $(G_2, s_2, t_2, H_2)$ are equivalent if $|V(G_1)| = |V(G_2)|$. We show how to compose a sequence of inputs $(G_1, s_1, t_1, H_1), \ldots, (G_r, s_r, t_r, H_r)$ on $n$ vertices each. Assume the vertices of $V(G_i)$ are labeled $v_1, \ldots, v_n$ such that $v_1 = s_i$ and $v_n = t_i$. We build a graph $G^*$ with a vertex cover $X^*$, and a set of forbidden pairs $H^* \subseteq \binom{V(G^*)}{2}$ such that all forbidden pairs have at least one member in $X^*$.

1. For $j \in [n]$ add a vertex $v_j^*$ to $G^*$.

2. Add a special starting vertex $w$ to $G^*$.

3. For $i \in [r]$ add a vertex $z_i$ to $G^*$ and make it adjacent to $w$ and $v_1^*$.

4. For $1 \leq j < h \leq n$, add a *ladder* gadget $L_{j,h}$ with $n$ spokes to $G^*$:

   - Add vertices $t_{j,h}^i$ and $f_{j,h}^i$ to $G^*$ for $i \in [n]$.
   - Add the set of edges $\{\{t_{j,h}^i, t_{j,h}^{i+1}\}, \{t_{j,h}^i, f_{j,h}^{i+1}\}, \{f_{j,h}^i, f_{j,h}^{i+1}\}, \{f_{j,h}^i, t_{j,h}^{i+1}\} \mid i \in [n-1]\}$ to form the inside of the ladder.
   - Make $v_j^*$ adjacent to $t_{j,h}^1$ and $f_{j,h}^1$, and make $v_h^*$ adjacent to $t_{j,h}^n$ and $f_{j,h}^n$.

5. Repeat the following for each $i \in [r]$:

   - For $1 \leq j < h \leq n$, if $\{v_j, v_h\} \notin E(G_i)$ then add $\{\{z_i, x\} \mid x \in L_{j,h}\}$ to the set of forbidden pairs. Here $L_{j,h}$ denotes the set of $2n$ vertices on the ladder for $\{j, h\}$.
   - For $1 \leq j < h \leq n$, if $\{v_j, v_h\} \in E(G_i)$ then do as follows for all $q \in [n]$:
     - If $\{v_j, v_q\} \in H_i$ or $\{v_h, v_q\} \in H_i$ then add $\{z_i, f_{j,h}^q\}$ to $H^*$.

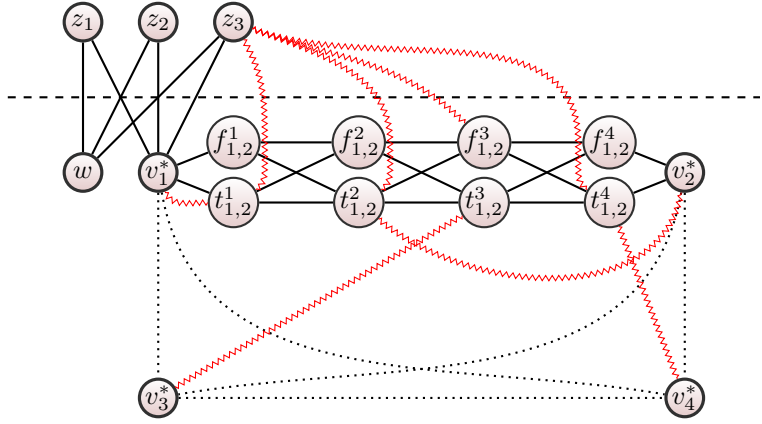

Figure 3: An example of the lower-bound construction of Theorem 11, cross-composing three instances with $n = 4$ into one. For clarity, only the vertices of the ladder $L_{1,2}$ are drawn; the other ladders are visualized by dotted paths. The forbidden pairs are drawn as zigzagged lines. Forbidden pairs with one element in $\{z_1, z_2\}$ are not drawn. For this example, the third input $(G_3, s_3, t_3, H_3)$ has a forbidden pair $\{v_1, v_3\} \in H_3$ causing the forbidden pair $\{z_3, f^3_{1,2}\} \in H^*$. The vertices below the horizontal dashed line form the vertex cover X*.

- Else, if $\{v_j, v_q\} \notin H_i$ and $\{v_h, v_q\} \notin H_i$, then add $\{z_i, t^q_{j,h}\}$.
- For $1 \leq j < h \leq n$, for $q \in [n]$, add $\{t^q_{j,h}, v^*_q\}$ to $H^*$.

6. This concludes the description of $G^*$ and $H^*$, which is illustrated in Fig. 3. Define $X^* := V(G^*) \setminus \{z_i \mid i \in [r]\}$.

It is easy to see that this construction can be carried out in polynomial time. The set $X^*$ is a vertex cover for $G^* \cup H^*$ since all edges and forbidden pairs of $G^*$ have at least one element in $X^*$. The size of $X^*$ is $1 + n + 2n\binom{n}{2}$, which is polynomial in $n$ and therefore the parameter $\ell := |X^*|$ of the constructed instance is suitably small. We set $s^* := w$ and $t^* := v^*_n$. Let us establish some properties of the constructed instance.

**Claim 3.** *Consider an $s^* - t^*$ path $P$ in $G^*$ which respects the forbidden pairs $H^*$, and orient the path such that it starts at $s^*$. The following must hold.*

1. *There is exactly one index $i^* \in [r]$ such that $z_{i^*} \in P$, and the path $P$ has the form $(s^* = w, z_{i^*}, v^*_1, \ldots, v^*_n = t^*)$.*

2. *If $P$ contains $v^*_j$ and $v^*_h$ then $\{v_j, v_h\} \notin H_{i^*}$, where $i^*$ is as defined above.*

3. *If $v^*_j$ and $v^*_h$ are vertices on $P$, and no vertices of the form $v^*_f$ are visited by $P$ between visiting $v^*_j$ and $v^*_h$, then $\{v_j, v_h\} \in E(G_{i^*})$, where $i^*$ is as defined above.*

*Proof.* (1) Since all vertices $z_i$ for $i \in [r]$ have the same neighborhood consisting of $w$ and $v^*_1$, if a path contains at least two such vertices then at least one of them is an endpoint of the path, which is not possible since $P$ is an $s^* - t^*$ path. Any $s^* - t^*$ path must use at least one vertex $z_{i^*}$ since all neighbors of $s^* = w$ have this form.

(2) Suppose $P$ contains $v^*_j$ and $v^*_h$, and assume for a contradiction that $\{v_j, v_h\} \in H_{i^*}$. Assume without loss of generality that $j < h$ and therefore $v^*_j \neq v^*_n$. Since $v^*_j$ is not an endpoint of $P$, vertex $v_j^*$ must have two neighbors on $P$. By construction of $G^*$ it easily follows that at least one of these neighbors must lie on a ladder. Assume that $v_j^*$ has a neighbor on a ladder $L_{j,p}$; the case that the neighbor lies on a ladder $L_{p,j}$ is symmetric. The forbidden pairs involving $z_{i^*}$ and the vertices of the ladder $L_{j,p}$ ensure that for each spoke of the ladder on vertices $\{t_{j,p}^q, f_{j,p}^q\}$ there is a forbidden pair containing $z_{i^*}$ and exactly one vertex of the pair. Since $P$ contains $z_{i^*}$ it must avoid exactly one vertex of each spoke of the ladder, which implies by construction of $G^*$ that all vertices on the ladder have only one valid option along which to continue the path, implying that $P$ must traverse the entire ladder to the other endpoint $v_p^*$. Since $\{v_j, v_h\} \in H_{i^*}$ the definition of $G^*$ shows that $\{z_{i^*}, f_{j,p}^h\} \in H^*$, and therefore path $P$ must visit the other vertex $t_{j,p}^h$ of the spoke when traversing the ladder. But by the last step of the construction we know that $\{t_{j,p}^h, v_h^*\} \in H^*$ is a forbidden pair of which $P$ contains both vertices; a contradiction.

(3) Suppose that $P$ successively visits the $v^*$-vertices $v_j^*$ and $v_h^*$ with $j < h$. By construction of $G^*$ it is easy to see that $P$ must traverse the vertices of the ladder $L_{j,h}$. Since $P$ contains $z_{i^*}$, and $z_{i^*}$ forms a forbidden pair with every vertex on the ladder $L_{j,h}$ if $\{v_j, v_h\} \notin E(G_{i^*})$, it follows that if $P$ can traverse the ladder $L_{j,h}$ without hitting a forbidden pair then the corresponding edge must be contained in $G_{i^*}$. □

With these structural properties of the constructed instance we can prove the correctness of the cross-composition. Assume that $G^*$ is a YES-instance, and let $P$ be an $s^* - t^*$ path in $G^*$. By (1) path $P$ has the form $(w, z_{i^*}, v_1^*, \ldots, v_n^*)$. Orient $P$ from $s^*$ to $t^*$, and consider the vertices $(v_{i_1}^*, \ldots, v_{i_m}^*)$ in the set $\{v_i^* \mid i \in [n]\}$ which are successively visited on this path. By (2) it follows that there are no indices $i_j, i_f$ such that $\{v_{i_j}, v_{i_f}\} \in H_{i^*}$, and by (3) the set $E(G_{i^*})$ contains all edges $\{v_{i_j}, v_{i_{j+1}}\}$ for $j \in [m-1]$. Hence the vertices $(v_{i_1}, \ldots, v_{i_m})$ form a path in $G_{i^*}$ which does not contain a forbidden pair. By the form of $P$ described in (1) it follows that $v_{i_1}^* = v_1^*$, and since $P$ ends with $v_n^*$ it follows that $v_{i_m}^* = v_n^*$. Hence this suggested path is a $v_1 - v_n$ path in $G_{i^*}$ which avoids forbidden pairs, which proves that instance number $i^*$ is YES.

For the reverse direction, suppose that $G_{i^*}$ contains a path $(v_{i_1}, \ldots, v_{i_m})$ with $i_1 = 1$ and $i_m = n$ which does not contain a forbidden pair. Now construct a path in $G^*$ which starts at $s^* = w$ and successively visits $z_{i^*}$ and $v_1^*$. It then traverses the ladder to $v_{i_2}^*$, and by construction of $G^*$ there is exactly one vertex on each spoke of the ladder which is not in a forbidden pair with $z_{i^*}$; the path uses these vertices. We can continue this, alternatingly traversing a vertex $v_{i_j}^*$ and a ladder, until reaching $v_{i_m}^* = t^*$. It is not difficult to verify that the constructed path does not contain any forbidden pairs of $H^*$, which concludes the proof. □

It is easy to modify the construction of Theorem 11 to prove a kernel lower bound for LONGEST PATH WITH FORBIDDEN PAIRS Parameterized by a Vertex Cover of $G \cup H$. The definition of the latter problem does not allow us to specify the endpoints $s$ and $t$ of the path, but by creating two suitably long paths and connecting these only to $s$ and $t$ we can ensure that a sufficiently long path in the resulting graph is actually an $s - t$ path. Also, similarly as before the hardness results for the existence problem of $s - t$ paths immediately carry over to long and short $s - t$ paths. We obtain the following results as a corollary.

**Corollary 6.** *The problems* SHORTEST $s - t$ PATH WITH FORBIDDEN PAIRS, LONGEST $s - t$ PATH WITH FORBIDDEN PAIRS, *and* LONGEST PATH WITH FORBIDDEN PAIRS, *do not admit a polynomial kernel parameterized by a vertex cover of $G \cup H$ unless $NP \subseteq coNP/poly$.*

Table 1 contains a summary of the results of this section.

|  | vc($G$) | vc($H$) | tw($H$) | tw($G \cup H$) | vc($G \cup H$) |
|---|---|---|---|---|---|
| $s-t$ Path F.P. | W[1]-hard | FPT | Para-NP-c | FPT | No poly |
| Shortest $s-t$ Path F.P. | W[1]-hard | FPT | Para-NP-c | FPT | No poly |
| Longest $s-t$ Path F.P. | W[1]-hard | Para-NP-c | Para-NP-c | FPT | No poly |
| Longest Path F.P. | W[1]-hard | Para-NP-c | Para-NP-c | FPT | No poly |

Table 1: Complexity overview of path problems with forbidden pairs. Each column represents a different parameterization; vc($G$) denotes the minimum vertex cover size of $G$, and tw($G$) denotes its treewidth. F.P. abbreviates "with Forbidden Pairs". The classification "No poly" means "no polynomial kernel unless NP $\subseteq$ coNP/poly", and "Para-NP-c" means "NP-complete for a constant value of the parameter". For a parameterization in FPT, we either list "FPT" or "No poly", depending on which of the two is more relevant: all problems listed as "No poly" are in FPT, and no problem listed as "FPT" admits a polynomial kernel. Shortest Path F.P. is trivially in $P$.

## 7 Conclusion

In this work we have shown that for sufficiently strong structural parameterizations, many path and cycle problems admit polynomial kernels even though their natural parameterizations do not. The marking technique using bipartite matching yields quadratic-vertex kernels for many problems parameterized by the size of a vertex cover. We introduced a binary encoding trick which gives polynomial kernels for problems parameterized by the max leaf number or by a modulator to cographs. On the negative side, we also exhibited smaller structural parameters which provably do not lead to polynomial kernels for Hamiltonian Cycle unless NP $\subseteq$ coNP/poly. Let us reflect briefly on the parameters used for the upper- and lower bounds.

Recall that the vertex cover number of a graph can also be interpreted as the number of vertex-deletions needed to reduce the graph to an independent set, i.e. the vertex-deletion distance to a graph of treewidth 0. Hence Theorem 3 shows that Long Cycle admits a polynomial kernel parameterized by vertex-deletion distance to treewidth 0. On the other hand, Theorem 8 shows that if NP $\not\subseteq$ coNP/poly then Hamiltonian Cycle does *not* have a polynomial kernel parameterized by the deletion distance to treewidth two (since outerplanar graphs have treewidth at most two), and of course this carries over to the harder problem Long Cycle. It is interesting to settle what happens for treewidth one, i.e., forests: does Hamiltonian Cycle parameterized by a feedback vertex set admit a polynomial kernel? To generalize the result of Theorem 6 by distance to a cluster graph, one could consider the distance to cographs.

The kernelization complexity of compound parameterizations remains largely unexplored: for example, how does the Long Cycle problem behave when parameterized by the solution size plus the vertex-deletion distance to an outerplanar graph? It follows from the work of Bodlaender, Thomassé, and Yeo [7] that Disjoint Paths and Disjoint Cycles do not admit polynomial kernels parameterized by the target value $k$ plus the deletion distance to a path. We hope that a search for polynomial kernels of structural parameterizations leads to reduction rules which are useful in practice.